 \documentclass[twocolumn]{aastex62}

\usepackage{color}

\shorttitle{Novae without Mass-Loss} 
\shortauthors{Kato et al.}

\begin{document}


\title{ASASSN-16\lowercase{oh}: A nova outburst with no mass ejection 
-- A new type of supersoft X-ray source in old populations}


\author{Mariko Kato} 
\affil{Department of Astronomy, Keio University, Hiyoshi, Yokohama
  223-8521, Japan}
 \email{mariko.kato@hc.st.keio.ac.jp}


\author{Hideyuki Saio}
\affil{Astronomical Institute, Graduate School of Science,
    Tohoku University, Sendai 980-8578, Japan}

\author{Izumi Hachisu}
\affil{Department of Earth Science and Astronomy, College of Arts and
Sciences,  University of Tokyo, 3-8-1 Komaba, Meguro-ku,
Tokyo 153-8902, Japan}



\begin{abstract} 
ASASSN-16oh is a peculiar transient supersoft X-ray source 
without a mass ejection signature in the field of the Small Magellanic 
Cloud. Maccarone et al. (2019) concluded that ASASSN-16oh is the first 
dwarf nova with supersoft X-ray that originated from an equatorial
accretion belt on a white dwarf (WD). Hillman et al. (2019) proposed 
a thermonuclear runaway model that both the X-rays and $V$/$I$ photons 
are emitted from the hot WD. 
We calculated the same parameter models as Hillman et al.'s 
and found that they 
manipulated on/off the mass-accretion, and their best fit $V$ light curves
are 6 mag fainter, and decay about 10 times slower, than that of ASASSN-16oh. 
We propose a nova model induced by a high rate of mass-accretion during 
a dwarf nova outburst, i.e., the X-rays originate from the surface 
of the hydrogen-burning WD whereas the $V/I$ photons are
from the irradiated disk. Our model explains the main 
observational properties of ASASSN-16oh. 
We also obtained thermonuclear runaway models with no mass ejection 
for a wide range of parameters of the WD mass and mass-accretion rates 
including both natural and forced novae in low-metal environments 
of $Z=0.001$ and $Z=0.0001$. 
They are a new type of periodic supersoft X-ray sources with no mass ejection, 
and also a bright transient in $V$/$I$ bands if they have a large disk.  
We suggest that such objects are candidates of Type Ia supernova
progenitors because its mass is increasing at a very high efficiency
$(\sim 100 \%)$.  
\end{abstract}


\section{Introduction}
\label{sec_introduction}

A nova is a thermonuclear runaway 
event on a mass-accreting white dwarf (WD).  
Novae usually brighten up by $\Delta V\sim 10$ mag or more in a few or tens 
days and decline in a few months or years.
After hydrogen ignites on the WD, the hydrogen-rich envelope expands
to giant size and emits strong winds \citep[e.g.][]{kat94h}.  
The nova brightness is dominated by free-free emission from 
optically thin ejecta outside the photosphere.   
Thus, the peak optical brightness depends on the 
maximum wind mass-loss rate \citep[e.g.,][]{hac06kb, hac15k}. Due to 
strong mass loss, the envelope loses its mass and the photosphere 
gradually shrinks and the photospheric temperature increases.  
Eventually the nova enters a supersoft X-ray source (SSS) phase. 
The envelope mass decreases further due to nuclear burning 
after the winds stop.  
Hydrogen burning ends when the envelope mass reaches a critical value.
As a definition, a nova accompanies strong winds (mass-ejection) 
in the optical bright phase.  

On the other hand, a dwarf nova is an accretion event on a compact object,
in which the mass accretion is enhanced by a thermal instability of an
accretion disk.  Dwarf novae usually brighten up by $\Delta V\sim 2-9$ mag 
in a day or so and stays at the brightness from a few days to a few weeks.
However, no SSS phase has ever been observed in dwarf nova outbursts. 

ASASSN-16oh is a peculiar transient SSS in the field of the Small
Magellanic Cloud (SMC), which was discovered by the All Sky Automated
Survey for Supernovae (ASASSN) on UT 2.15 December 2016 (JD~2,457,724.65)
at $V= 16.9$ \citep{jha16}.  The brightness reaches an absolute 
magnitude of $M_V= -2.3$
on JD~2,457,744.6, where we assume that the distance modulus is 
$\mu_0\equiv (m-M)_0=18.9$ and the absorption in the $V$ band is $A_V= 0.13$
toward ASASSN-16oh \citep{jha16, mro16}.  
The Optical Gravitational 
Lensing Experiment IV (OGLE-IV) \citep{uda15} data show
that the object is an irregular variable for several years 
with the quiescent luminosity of $I=20.3$ and $V=21.1$ \citep{mac19}.

The $V$ and $I$ light curves of ASASSN-16oh show a resemblance 
with those of long orbital-period dwarf novae such as V1017~Sgr 
in the peak brightness, outburst amplitude, and timescales 
(see Section \ref{sec_obs} for detail). 
The orbital period of V1017~Sgr is $P_{\rm orb}= 5.786$~days
\citep{salazar17}.  Therefore, \citet{mac19} expected the orbital 
period of ASASSN-16oh to be several days.           

\citet{mac19} concluded that ASASSN-16oh is not a classical nova 
(thermonuclear runaway event) but an accretion event like a dwarf nova,
mainly because (1) the optical spectra show a very narrow width of
\ion{He}{2} emission with no signature of mass-ejection, (2) very slow 
rise ($\sim 200$ days) to the optical maximum compared with those of
classical novae (a few days), (3) rather dark peak $V$ 
magnitude of $M_{V, {\rm max}} \sim -2.3$ compared with those of
recurrent novae \citep[e.g., the 1 year recurrence period nova
M31N~2008-12a of $M_{V, {\rm max}}= -7.2$,][]{dar15, hen18}.

Only the concern is the origin of supersoft X-rays because  
no SSS phase has been detected in dwarf nova outbursts.
\citet{mac19} interpreted the supersoft X-rays could originate from 
a spreading layer around the equatorial accretion belt. 
This model requires a very high mass accretion rate of 
$\sim 3\times 10^{-7}~M_\sun$~yr$^{-1}$ 
($2 \times 10^{19}$~g~s$^{-1}$) on a very massive WD of $1.3~M_\sun$ 
to support high temperatures and fluxes for supersoft X-ray. 

The hot spreading layer of the accretion belt has not yet been 
studied in detail. If a disk around a massive WD 
with high accretion rates always accompany a hot spreading layer, 
such supersoft X-rays could be observed 
in recurrent novae in their quiescent phase in U Sco 
\citep[$M_{\rm WD} \sim 1.37~M_\sun$ with $\dot M_{\rm acc} 
\sim 2.5 \times 10^{-7}~M_\sun$yr$^{-1}$,][]{hkkm00} and RS Oph
\citep[$\sim 1.35~M_\sun$ with $\sim 2 \times 10^{-7}~M_\sun$
yr$^{-1}$,][]{hac08kl},  
but not yet detected (see Section \ref{sec_rsoph})
This gives us the idea that the supersoft X-rays could originate 
from thermonuclear flashes on the WD.

It should be noted that the high instantaneous mass accretion rate 
($\sim 3\times 10^{-7}~M_\sun$~yr$^{-1}$) \citep{mac19} 
is close to the critical value to maintain a steady hydrogen shell-burning, 
otherwise very weak shell flashes \citep[see, e.g., Figure 6 of][]{kat14shn}.
With such a high rate mass-accretion, nova outbursts are so weak that
wind mass-loss is also weak. Moreover, a low-metallicity (low-$Z$)
environment in SMC suggests weak mass ejection during nova outbursts.
These condition led us to search nova (thermonuclear runaway event)
solutions for no-mass-ejection \citep[e.g.][]{kat85, kat09}. 
They are periodically variable supersoft X-ray sources 
\citep[e.g.,][]{hac16}. 
Such an object has not been observationally detected yet.

\begin{figure*}
\plotone{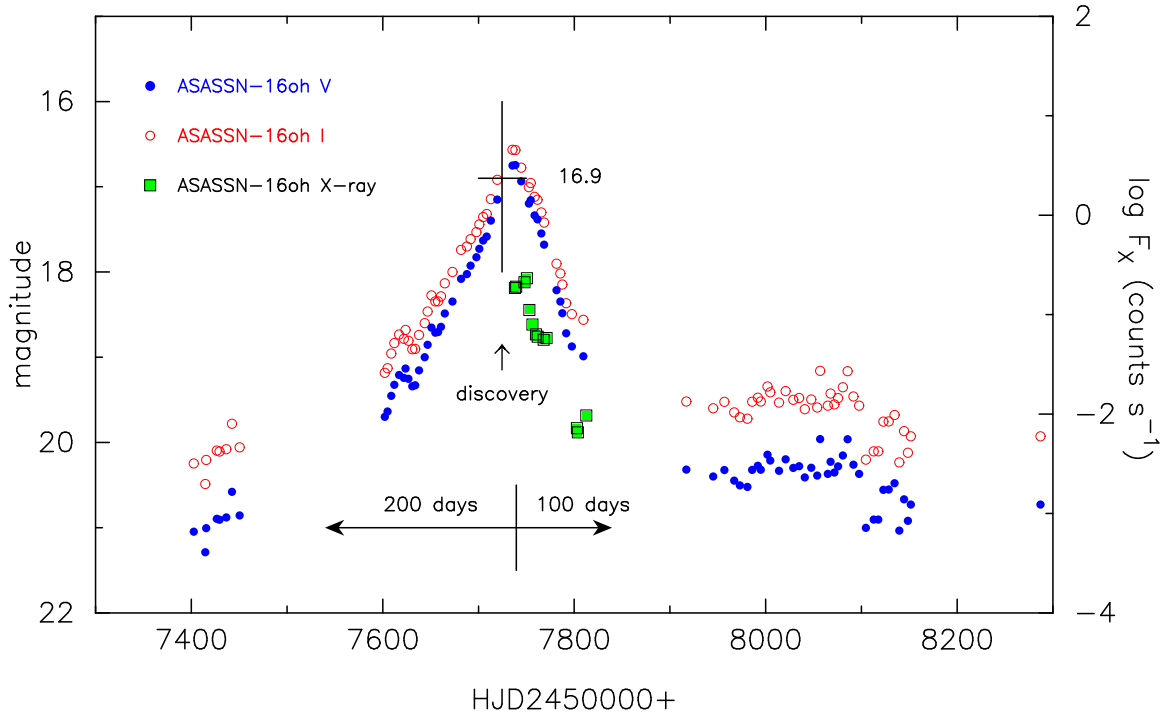}
\caption{
The $I$ and $V$ light curves of ASASSN-16oh as well as the soft X-ray 
count rates observed with {\it Swift}.  The data are taken from
the OGLE~IV and {\it Swift} web-sites.  The $V$ magnitudes are
estimated from the $I$ data together with several $V-I$ colors
in the outburst (see the main text for more details).
\label{as16oh_vi_linear}}
\end{figure*}

\citet{hil19} proposed another interpretation of ASASSN-16oh, 
a thermonuclear runaway event, and calculated several shell 
flash models. 
They claimed that X-rays and $V/I$ photons are emitted 
from a WD that undergoes hydrogen burning. However, 
it is unlikely that a hot WD with a surface temperature $T > 400,000$~K
emits many low energy $V/I$ photons as well as X-rays. 
This concern led us to start the present work.
We have calculated a nova model with the same parameters as 
Hillman et al.'s model, and obtained different results. 
This will be presented in Section \ref{hillmanmodel}.

It is reasonable that X-rays and optical $V/I$ photons come from 
different places in the binary system as interpreted by \citet{mac19}. 
We regard that the $V/I$ photons originated from a large accretion 
disk irradiated by the hot WD. 
We propose a recurrent nova (thermonuclear runaway)
model for the supersoft X-rays and search parameter region for
no-mass-ejection novae. 
The main differences from Hillman et al.'s model are:
(1) The $V/I$ photons originate from the accretion disk and 
companion star because the WD is too hot to emit many $V/I$ photons.  
(2) To explain the observed X-ray light curve, the WD mass is 
much higher ($1.32~M_\sun$) than that of Hillman et al.'s model 
($1.1~M_\sun$). 
(3) We suggest that the hydrogen shell flash could be 
triggered by a massive mass-inflow during a dwarf nova outburst. 

Both models of dwarf nova \citep{mac19}
and thermonuclear shell flash (present work) are based on the presence of 
a mass-accreting massive WD ($M_{\rm WD} \ga 1.3~M_\sun$) in ASASSN-16oh. 
In the latter model, the WD mass increases steadily 
because of no mass-ejection during the flash. 
Such a massive mass-increasing WD in a low $Z$ environment is 
very interesting in binary evolution scenarios because it could be 
a new kind of candidates for Type Ia supernova progenitors.

The aim of this work is to present a theoretical model 
of ASASSN-16oh as well as to search for the possible region of 
novae with no mass ejection in low metallicity environments. 
We organize the present paper as follows.  
First, we summarize the observational properties of ASASSN-16oh 
in Section \ref{sec_obs}. 
Then, we briefly introduce possible parameter region 
for nova outbursts without mass-ejection 
including both forced and natural novae in Section \ref{model}. 
Our results are presented in Sections \ref{z001}. 
We examine Hillman et al.'s (2019) nova model in 
Section \ref{hillmanmodel}, 
and compare our model with 
Maccarone et al.'s model in Section \ref{sec_maccarone}. 
Discussion and conclusions follow in Sections \ref{discussion} and 
\ref{conclusion}, respectively.

\section{Observational properties of ASASSN-16\lowercase{oh}} \label{sec_obs}

In this section we summarize the characteristic properties of ASASSN-16oh.  

\subsection{$V/I$ Light Curves}
We plot the $I$ magnitudes of ASASSN-16oh in Figure \ref{as16oh_vi_linear} 
(unfilled red circles), taken from the OGLE-IV website \citep{mro16, mac19}.  

For later use, we have estimated the $V$ light curve from the $I$ data 
and several $V-I$ colors listed in \citet{mac19}; 
$V-I= 0.8$ in quiescent phase, $V-I= 0.36$
on JD 2457657.7, $V-I= 0.34$ on JD 2457681.78, and $V-I= 0.16$
on JD 2457744.6.  With a linear interpolation of these $V-I$ data
and assumed (start, end) of the outburst to be 
(200 days before, 100 days after) the $I$ peak (JD~2457737.1), 
we recover the $V$ magnitudes (filled blue circles)
as shown in Figure \ref{as16oh_vi_linear}.
We assumed that the $I$ brightness reaches maximum on
JD~2457737.1 from an interpolation of the other $I$ data. 

The characteristic property of the light curve is a slow 
rise, sharp peak followed by a rapid decay. 
The brightness rises slowly in $\sim 200$ days and decline fast
in $\sim 100$ days. This properties are unlike to 
typical nova outbursts but rather similar to long orbital-period 
dwarf novae. 
For example, the dwarf nova outbursts of V1017~Sgr
show a 90 days rise and 90 days decline 
with almost symmetric light curve shape. 
In contrast, a typical nova rises within a few days and decays 
much slowly in a timescale of years \citep[see, e.g., Figure 1 of][]{hac15k}.

There are other common properties between ASASSN-16oh and the dwarf nova
V1017~Sgr. 
The peak $V$ brightness is $M_{V, {\rm max}} \sim -2.3$ in ASASSN-16oh,  
similar to $M_{V, {\rm max}} \sim -1.3$ in V1017~Sgr \citep{schaefer18}. 
The quiescent brightnesses of the both objects are also similar, that is,
$M_{V, {\rm min}}\approx 2.1$ for ASASSN-16oh and
$M_{V, {\rm min}}\approx 1.8$ for V1017~Sgr \citep{schaefer18}. 
Thus, the light curve shape of ASASSN-16oh may be explained as 
a long orbital-period dwarf nova.

\subsection{Optical Spectrum}

\citet{mac19} presented the Southern African Large Telescope (SALT)
optical spectrum of ASASSN-16oh 
taken on UT 2016 December 16--24 (JD~2457738.5--245746.5),   
just after the $I$ peak (JD~2457737.1). This spectrum shows 
no indication of mass loss, such as many broad emission 
lines or P-Cygni profiles usually seen in classical novae. 
Thus, if this is a classical nova phenomenon, it is a rare case in which
no wind mass-loss occurs.

\subsection{X-ray Light Curve and Spectrum}

A distinct difference between ASASSN-16oh and dwarf novae is the supersoft
X-ray detection.  No SSS phase has ever been detected in dwarf novae
while a SSS phase is frequently observed in classical novae
\citep{schw11}.  If ASASSN-16oh is a dwarf nova, 
this is the first case of supersoft X-ray detection \citep{mac19}. 

Figure \ref{as16oh_vi_linear} also shows the X-ray count rates
observed with {\it Swift} 
($0.3-10.0$ keV: filled green squares outlined with black line).  
It seems that the X-ray count rate reaches maximum slightly 
later than the $I$ peak and decays with the $I$ light curve. 

The X-ray spectra are dominated with soft components. 
\citet{mac19} deduced a blackbody temperature of 900,000~K 
and a flux of $\sim 1\times 10^{37}$ erg~s$^{-1}$ 
from the {\it Swift}/XRT spectrum on UT 15 December 2016
(JD~2,457,737.5) at the distance of the SMC. 
From the {\it Chandra}/LETG spectrum taken on 
UT 28 December 2016 (JD~2,457,750.5), \citet{mac19} 
derived a blackbody temperature of 905,000~K,
a luminosity of $6.7 \times 10^{36}$ erg~s$^{-1}$, and
the hydrogen column density of $N_{\rm H}= 3.4 \times 10^{20}$~cm$^{-2}$,   
and, for an atmosphere model of solar abundance, 
750,000~K and $N_{\rm H}= 2.0 \times 10^{20}$~cm$^{-2}$.  
\citet{hil19} re-analyzed the same {\it Chandra} spectrum obtained
by \citet{mac19} with a metal-poor atmosphere model and obtained
the effective temperature of 750,000~K, the bolometric luminosity of 
$4.3 \times 10^{36}$ erg~s$^{-1}$, 
and the hydrogen column density of $N_{\rm H}= 2.3 \times 10^{20}$~cm$^{-2}$.

These temperatures are consistent with the surface temperatures of 
massive WDs in the SSS phase of classical novae \citep[e.g.,][]{kat97,schw11}. 
Thus, nuclear burning origin seems to be a natural explanation of the X-rays.

\subsection{Column Density and Unabsorbed X-ray Flux}

Both \citet{mac19} and \citet{hil19} obtained a similar column density 
of $N_{\rm H}= 2 \times 10^{20}$~cm$^{-2}$  that can be converted to
$E(B-V)= N_{\rm H}/ 5.8 \times 10^{21} \approx 0.035$ \citep{boh78},
$E(B-V)= N_{\rm H}/ 6.8 \times 10^{21} \approx 0.03$ \citep{guv09},
or $E(B-V)= N_{\rm H}/ 8.3 \times 10^{21} \approx 0.024$ \citep{lis14}.
In the present paper, we adopt $A_V= 3.1 E(B-V)= 0.1$, 
$(m-M)_V= 18.9 + 0.1=19.0$, and $(m-M)_I= 18.95$.

\citet{mac19} obtained the unabsorbed X-ray luminosity to be 
$\sim 1\times 10^{37}$ erg~s$^{-1}$ from the {\it Swift}/XRT spectrum  and 
to be $6.7 \times 10^{36}$ erg~s$^{-1}$ from the {\it Chandra}/LETG spectrum.
\citet{hil19} obtained the unabsorbed X-ray flux 
to be $4.3\times 10^{36}$~erg~s$^{-1}$. 

These luminosities are much lower than the Eddington luminosity 
$\sim 1 \times 10^{38}$~erg~s$^{-1}$ for a $1.0~M_\sun$ WD.
If it is a classical nova outburst, 
the X-ray luminosity from the naked WD is close to the Eddington limit. 
\citet{hil19} suggested that the low brightness is owing 
to occultation by the accretion disk rim like in the recurrent 
nova U Sco, although they did not include effects of 
the accretion disk in their $V$ light curve model. 
\citet{ori13} estimated the unabsorbed X-ray luminosity 
of U~Sco in the SSS phase to be $\sim 7\times 10^{36}$~erg~s$^{-1}$ 
from the {\it Chandra}/HRC-S and LETG spectra.  The surface 
of the WD is occulted by the disk rim and the X-rays come 
from Thomson scattering by expanded plasma around the WD.
This X-ray luminosity is similar to that of ASASSN-16oh.

\begin{figure}
\epsscale{1.1}
\plotone{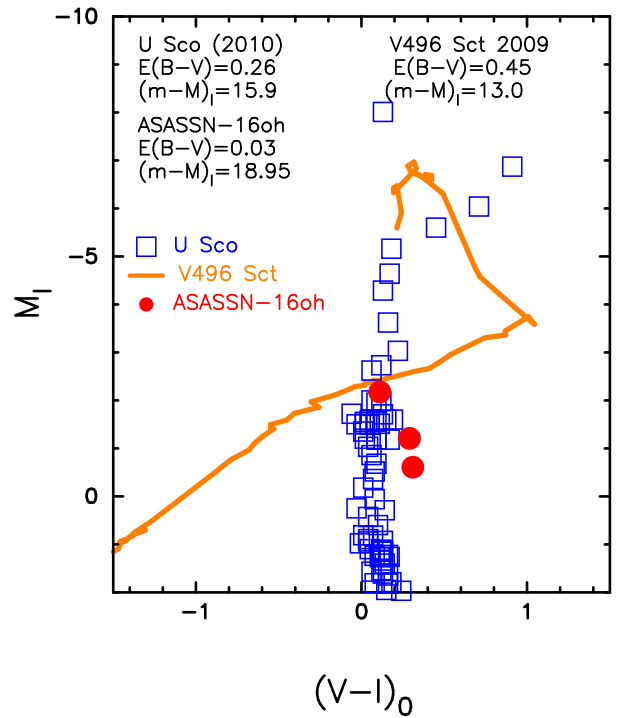}
\caption{The $(V-I)_0$-$M_I$ color-magnitude diagram of ASASSN-16oh 
(filled red circles) in outburst.  Here, $(V-I)_0$ is the intrinsic $V-I$ color
and $M_I$ is the absolute $I$ magnitude and $(V-I)_0$ is calculated from
the relation of $(V-I)_0= (V-I) - 1.6 E(B-V)$ \citep{rie85}. 
Three $V-I$ colors of ASASSN-16oh are $V-I= 0.36$ (JD 2457657.7), 
$V-I= 0.34$ (JD 2457681.78), and $V-I= 0.16$ (JD 2457744.6). 
For comparison, the color-magnitude tracks of the recurrent nova U~Sco
(open blue square) and the classical nova V496 Sct (orange line) are added.
The distance modulus and color excess of $(m-M)_I=15.9$ and $E(B-V)=0.26$
(U~Sco), and 13.0 and 0.45 (V496~Sct), are taken
from \citet{hac18kb} and \citet{hac19ka}, respectively.
\label{hr_diagram_asassn16oh_u_sco_outburst_vi}}
\end{figure}

\begin{figure*}
\plotone{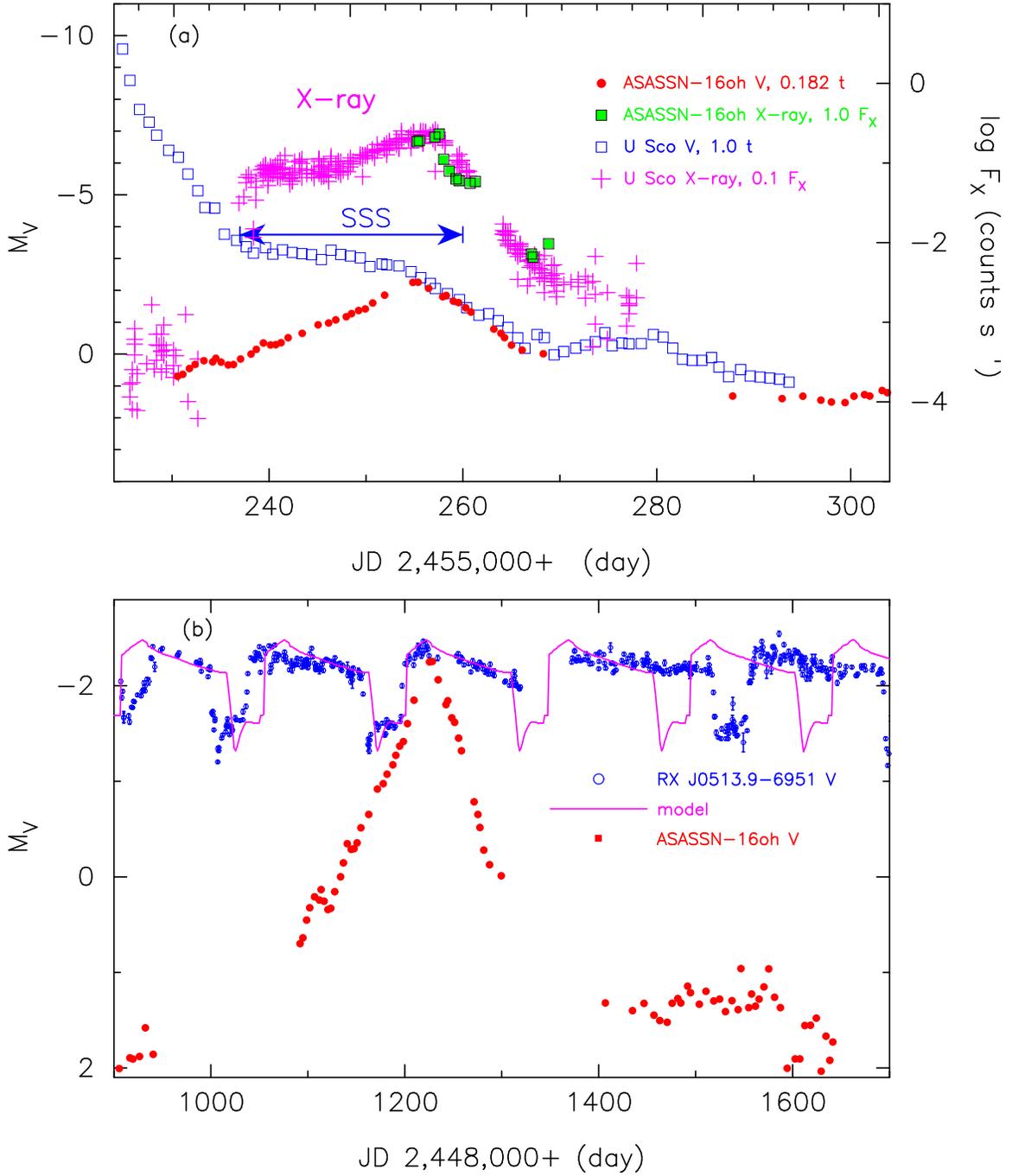}
\caption{(a) The $V$ and supersoft X-ray light curves of the recurrent 
nova U~Sco (2010) compared with the $V$ light curve of ASASSN-16oh.
The data of U~Sco are taken from \citet{pagnotta15}.
The X-ray data are taken from the {\it Swift} website \citep{eva09}.
The timescale of ASASSN-16oh is squeezed by 5.5 times.
The peak brightness of ASASSN-16oh is compatible with the brightness
of U~Sco in the SSS phase.  The $V$ brightness of U~Sco in the SSS phase
can be explained by a large accretion disk irradiated 
by a central hot WD \citep[see, e.g.,][]{hkkm00}.
(b) The $V$ light curve of the LMC supersoft X-ray source RX~J0513.9$-$6951
and a model light curve (solid magenta line) calculated by 
\citet{hac03RXJ}.  The data of RX~J0513.9$-$6951 are the same as those
in \citet{hac03RXJ}.  We add the $V$ light curve of ASASSN-16oh for 
comparison, assuming $(m-M)_V=19.0$.
\label{vmagfit_u_sco_rxj0513_as16oh}}
\end{figure*}

\subsection{$V-I$ Color: Evidence 1 of an Irradiated Accretion Disk}

Figure \ref{hr_diagram_asassn16oh_u_sco_outburst_vi}
shows the color-magnitude diagram, $(V-I)_0$-$M_I$,  
of ASASSN-16oh, around the peak of the outburst 
($t= -79.4$, $-55.3$, and $+7.5$ days after the $I$ peak).
Here, $(V-I)_0$ is the intrinsic $V-I$ color, which is calculated from
the relation of $(V-I)_0= (V-I) - 1.6 E(B-V)$ \citep{rie85},
and $M_I$ is the absolute $I$ magnitude.  

Figure \ref{hr_diagram_asassn16oh_u_sco_outburst_vi}
also shows the color-magnitude track of the recurrent nova U~Sco 
in the 2010 outburst.  We assume the distance modulus in the $I$ band
to be $(m-M)_I= 15.9$ and the reddening to be $E(B-V)= 0.26$
after \citet{hac18kb}.
The $UBVI_{\rm C}$ data of U~Sco are taken from \citet{pagnotta15}.  
Except for the short period after the peak ($M_I < -5$), the $(V-I)_0$
color of U~Sco is almost constant,  i.e., $(V-I)_0 \sim 0.1-0.2$. 
U Sco has an irradiated accretion disk with the radius of 
$\sim 2.5~R_\sun$ \citep{hkkm00}.  U Sco entered a plateau phase 
in the $V/I$ light curves when $I > 14$, that is, when $M_I \gtrsim -5$ 
as shown in Figure 2 of \citet{pagnotta15}. This is because 
the $V$ and $I$ brightnesses are dominated by the 
irradiated accretion disk \citep{hkkm00}.
This plateau phase corresponds almost to the SSS phase.  
The $(V-I)_0$ colors of ASASSN-16oh well overlap 
those of U~Sco in its SSS phase (when $M_I \gtrsim -2$). 

For comparison, we further add the track of a typical classical nova,
V496 Sct.  We assume the distance modulus in the $I$ band
to be $(m-M)_I= 13.0$ and the reddening to be $E(B-V)= 0.45$
after \citet{hac19ka}. 
The $BVI_{\rm C}$ data of V496~Sct are taken from the
archives of the Variable Star Observers League of Japan (VSOLJ),
the American Association of Variable Star Observers (AAVSO), 
the Small and Medium Aperture Telescope System \citep[SMARTS,][]{wal12},
and \citet{raj12}.  
The $V$ and $I$ light curves are plotted in Figure 52 of \citet{hac19ka} 
and the $(V-I)_0$-$M_I$ color-magnitude diagram is 
presented in \citet{hac19kb}.
In usual classical novae, their optical spectra are dominated by 
free-free emission from optically thin ejecta.  
After the optical peak, the optical brightness monotonically decreases
with time and its color goes toward the red due mainly to the emission
line effect in the $I$ band such as \ion{O}{1} and \ion{Ca}{2} triplet.
In the later phase ($M_I > -4$), the nova enters 
the nebular phase, and strong emission lines such as [\ion{O}{3}]
contribute to the $V$ band.  Then, the $V-I$ color turns to the blue. 
No X-ray data of V496 Sct are available, but the nova should enter
the SSS phase when it declined to $M_I > 0$ as shown in Figures
50 and 52 of \citet{hac19ka}.  

In general, classical novae are short orbital period binaries of 
$P_{\rm orb}\sim $ a few hours.  They have an accretion disk with
the radius of $\sim 0.5~R_\sun$, the brightness of which is rather
faint in the SSS phase.  Even if the accretion disk survives after
the nova outburst, it is deeply embedded in the ejecta. 
In the nebular phase, the accretion disk appears because the ejecta 
becomes optically thin but it is too small to contribute to 
the luminosity and color. 

In contrast, U~Sco has a large accretion disk with the radius of
$\sim 2.5~R_\sun$ because the orbital period is $P_{\rm orb}=1.23$ days
and its binary size (separation) is $a\sim 6-7 ~R_\sun$ \citep{hkkm00}.
Therefore, the irradiated disk substantially contributes to the $V$ and 
$I$ bands in the SSS phase \citep{hkkm00}.  This is partly because
the ejecta mass of U~Sco is very small and the nebular emission line effect
is relatively small.  
These physical properties give rise to the color-evolution difference
between U~Sco and V496 Sct. 

Because the $(V-I)_0$ color in the SSS phase is similar
between ASASSN-16oh and U~Sco, 
we expect that the $V$ and $I$ brightnesses of ASASSN-16oh originate
from the irradiated accretion disk like U~Sco.

\begin{figure*}
\epsscale{0.65}
\plotone{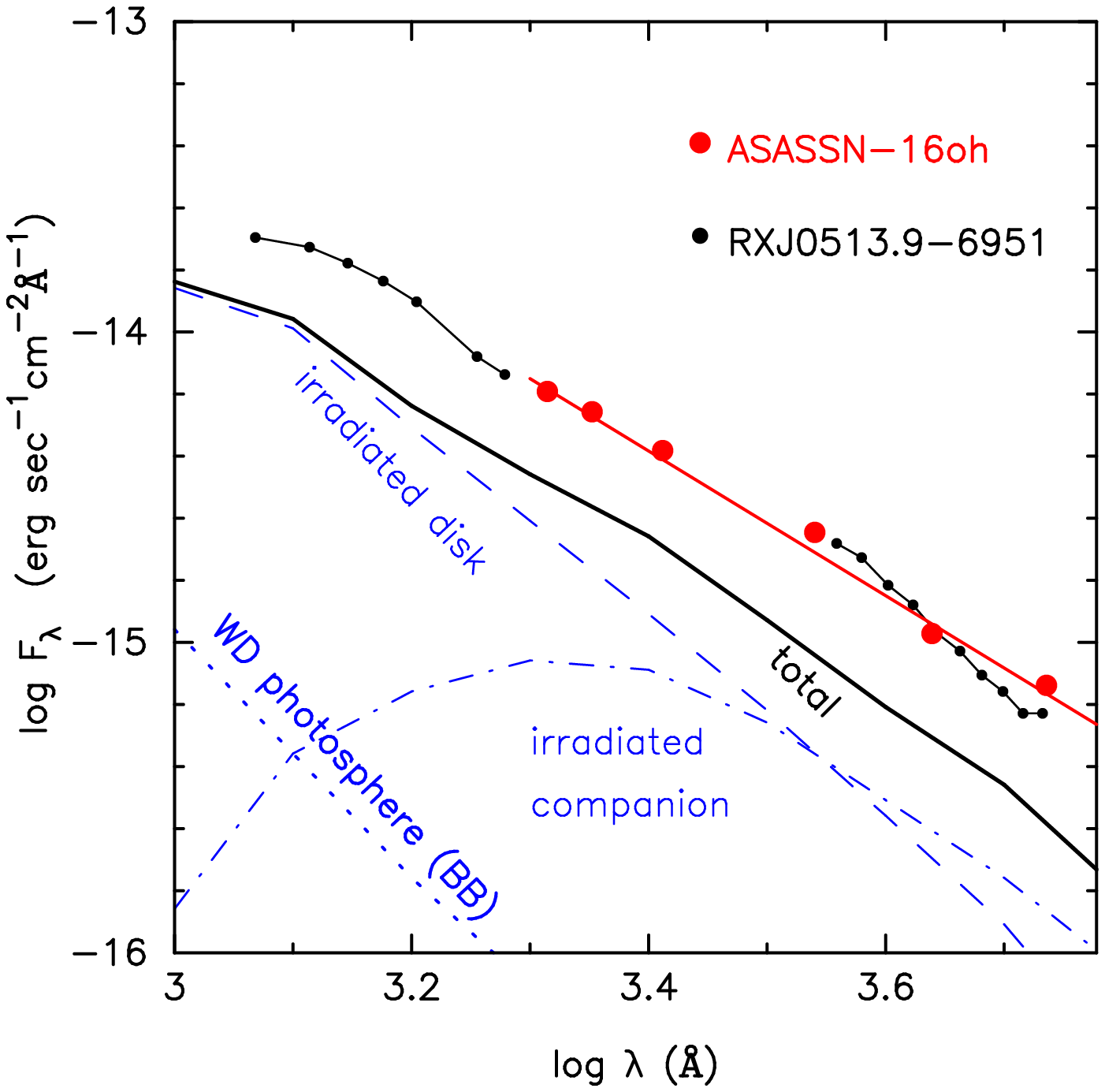}
\caption{ Spectral energy distribution of ASASSN-16oh (red dot: 
taken from \citet{mac19}) compared with RX J0513.9-6951 (black dot 
connected with solid line: \citet{pak93}).  The red line indicates 
the inclination of $F_\lambda \propto \lambda^{-2.33}$ law 
\citep{mac19}. The black and blue lines represent 
a theoretical model \citep[taken from ][]{pop96} 
for a supersoft X-ray source that consists of a steady burning WD 
(labeled BB, dotted line), irradiated accretion disk 
(dashed line), and irradiated companion star (dashed dotted line). 
The total flux (thick solid black line), which is the summation
of these three, shows similar spectrum energy distribution
to ASASSN-16oh and RX J0513.9-6951.  The theoretical lines and
fluxes of RX J0513.9-6951 are shifted downward by 
$\log$(60~kpc/50~kpc)$^2=0.158$ to compensate the distance difference
between the SMC and LMC.   
\label{uvot}}
\end{figure*}

\subsection{$M_V$ Light Curve: Evidence 2 of an Irradiated Accretion Disk}
\label{accretion_disk}

To further confirm the presence of an irradiated large accretion disk in 
ASASSN-16oh, we compare the absolute $V$ magnitude with two
other binaries that have a prominent irradiated accretion disk. 
One is the galactic recurrent nova U Sco, and the other is the Large
Magellanic Cloud (LMC) supersoft X-ray source RX~J0513.9$-$6951.

\subsubsection{U Sco}

Figure \ref{vmagfit_u_sco_rxj0513_as16oh}(a) shows the light 
curve of U~Sco in the 2010 outburst. We adopt the distance 
modulus in the $V$ band to be $(m-M)_V= 16.3$ \citep{hac18kb}.
Around the optical peak, the brightness is dominated 
by the emission from the ejecta. As time goes on the ejecta becomes 
optically thin and the continuum emission from the irradiated 
disk becomes dominant. In the SSS phase, the light curve shows 
a plateau because its brightness is dominated by
the irradiated accretion disk. The disk keeps a constant brightness until 
the nuclear burning turned off around JD~2,445,260.

We add the absolute $V$ magnitude, $M_{\rm V}$, light curve of
ASASSN-16oh to compare with that of U Sco. Here, the timescale of
ASASSN-16oh is squeezed by 5.5 times and the origin of time 
is shifted to match their supersoft X-ray peaks.

This figure demonstrates that the peak absolute magnitude $M_{\rm V}$ 
of ASASSN-16oh is roughly the same as that of U~Sco in the SSS phase. 
We expect that a larger irradiated disk is brighter. 
Thus, a rough agreement of the absolute brightness indicates that 
the sizes of the accretion disks are comparable.
The orbital period of U Sco is $P_{\rm orb}= 1.23$~days \citep{sch95r},
and the disk size is $\sim 2.5~R_\sun$ \citep{hkkm00}. 
This is broadly consistent with the orbital period (several days)
of ASASSN-16oh suggested by \citet{mac19}.

\subsubsection{RX~J0513.9$-$6951}

We also compare ASASSN-16oh with RX~J0513.9$-$6951.
This object shows a periodic variation in the $V$ magnitude as shown
in Figure \ref{vmagfit_u_sco_rxj0513_as16oh}(b). 
Here we adopt the distance modulus in the $V$ band 
of RX~J0513.9$-$6951 to be $(m-M)_V= \mu_0 + 3.1\times E(B-V)=
18.5 + 3.1\times 0.12= 18.9$.  We have $\mu_0= 18.5$ \citep{pie13}
and $E(B-V)= 0.12$ \citep{imara07} toward the LMC.
The supersoft X-ray flux is observed in the optical low state. 

\citet{hac03RXJ} modeled this object with a binary consisting of 
an accreting WD, large accretion disk, and lobe-filling companion star. 
The mass accretion rate $\dot M_{\rm acc} > 10^{-6}~M_\sun$ yr$^{-1}$ 
is above the stability line so that hydrogen is stably burning. 
Thus, the WD is always as bright as the Eddington luminosity.
At the beginning, the mass-transfer rate is larger than the critical 
rate of wind \citep[$\dot M_{\rm cr}$, see, e.g., Figure 6
of][]{kat14shn}.
The WD envelope expands to emit strong winds.
The winds hit the surface of the companion star and strip off
the very surface layer.  This effect works to reduce or stop 
the mass transfer.  The WD envelope shrinks and the winds stop.
The temperature of the WD photosphere increases and the main
emitting wavelength region moves to supersoft X-ray.
The very surface layer of the companion star recovers its original
radius in a thermal timescale and the mass transfer restarts.
The high rate mass accretion onto the WD is further delayed
by a viscous timescale of the accretion disk.  
After a total time of recovery (thermal timescale)
and diffusion (viscous timescale), the high rate of mass accretion 
recovers.  This is the mechanism of periodic variation of mass-transfer.  

In the wind phase, the surface of the accretion disk is blown in the
wind and the optically thick region of the accretion disk is enlarged
to beyond the size of the binary.  Because the irradiated surface
area becomes large, it brightens up.  This is the optical high state.
When the winds stop, the size of the accretion disk comes back 
to the original size.  Its brightness declines.  This is the optical
low state in the $V$ band.  Thus, the optical high state corresponds to the 
wind phase and expanded accretion disk. In a model of \citet{hac03RXJ}, 
the optically thick region of accretion disk expands comparable to the 
binary separation, $a=5.53~R_\sun$ for their binary model of 
$M_{\rm WD}= 1.3 ~M_\sun$ and $M_2= 2.6 ~M_\sun$.  The orbital period 
is $P_{\rm orb}= 0.76$~days \citep{pak93}.

We add the ASASSN-16oh data to Figure \ref{vmagfit_u_sco_rxj0513_as16oh}(b) 
in the same timescale. 
The peak $M_{V, \rm max}$ of ASASSN-16oh 
is almost the same as the optical high state of RX~J0513.9$-$6951. 
This is also an indication that ASASSN-16oh has 
a large irradiated disk comparable to that in RX~J0513.9$-$6951.

\subsection{UV/Optical SED: Evidence 3 of an Irradiated Accretion Disk}
\label{sec_sed}

\citet{mac19} obtained the spectral energy distribution (SED) of ASASSN-16oh 
from 2000\AA~ to 6000 \AA~ that is well approximated by a  
power law of $F_\lambda \propto \lambda^{-2.33}$ as in Figure \ref{uvot}.
Here, $F_\lambda$ is the energy flux at the wavelength $\lambda$.
For comparison, we added the continuum flux of the LMC supersoft X-ray
source RX J0513.9-6951, which is taken from the spectrum observed with
{\it the International Ultraviolet Explore (IUE)}/SWP and 
2.2 m Max-Planck ESO telescope \citep{pak93}.
This spectrum shows a resemblance to ASASSN-16oh, suggesting that 
the binary is consisting with a nuclear burning WD, irradiated 
accretion disk, and lobe-filling companion star like RX J0513.9-6951. 
We also add a model spectrum calculated by \citet{pop96}
for LMC supersoft X-ray sources (thick solid black line).
This model consists of a steady burning WD, irradiated accretion disk,
and lobe-filling companion star.  
The total flux, which is the summation of these three components, 
well reproduces the  $F_\lambda \propto \lambda^{-2.33}$ law. 
The irradiated accretion disk dominantly contributes to the total flux, 
whereas the irradiated secondary star contributes only to the optical region.
The contribution from the WD photosphere is negligible.  
The agreement in the SED between RX J0513.9-6951 and ASASSN-16oh 
also suggests the presence of an irradiated accretion disk in ASASSN-16oh.
We will discuss this point in more detail in Section \ref{sec_uvdisk}.

\section{Numerical Calculation of Hydrogen Shell Flashes with no-mass-ejection}
\label{model}

Based on the observational properties of ASASSN-16oh,  
it is natural to assume that X-rays and optical $V$/$I$ emission originate
from different places in the binary. We regard that 
the $V$/$I$ band emission comes from an irradiated accretion disk and
companion star, and the X-rays from a hot WD surface. 
Here, we focus our thought on shell flash models that show 
no-mass-ejection.

\subsection{Novae without Mass-Ejection}
As a definition, novae accompany a strong and fast mass ejection
(or strong and fast wind mass-loss) \citep[see, e.g.,][]{war95}.
The optically thick winds are accelerated when the photospheric luminosity 
approaches the Eddington luminosity. 
Because the OPAL opacity has a prominent peak 
at $\log T$ (K) $ \sim 5.2$ owing to iron ionization \citep{igl96},
the optically thick winds are accelerated around this temperature
region, i.e., deep inside the photosphere. 
The occurrence condition of winds was
examined first by \citet{kat85} for the old opacities, and then 
by \citet{kat09} and \citet{kat13hh} for the OPAL opacities \citep{igl96}. 
These works clarified that WDs do not emit optically-thick winds
in the following cases. 

\begin{enumerate}
\item
 In low mass WDs, the wind acceleration is insufficient so that
the wind velocity at the photosphere does not exceed the escape
velocity there.  The envelope simply expands and then shrinks
without mass-ejection.  This occurs in low mass WDs of
$M_{\rm WD}< 0.5-0.6~M_\sun$ for the metallicity of $Z=0.02$ \citep{kat09}.
The galactic slow nova PU Vul corresponds to this case \citep{kat11, kat12mh}. 

\item When the mass accretion rate is high and the ignition mass is
very small, the envelope does not expand to reach
$\log T$ (K) $ \sim 5.2$ at the photosphere.
Winds are not accelerated \citep{kat09}. 

\item In old population novae, the metallicity is very low so that
the iron peak in the OPAL opacity is not high enough to accelerate winds 
\citep{kat13hh}. 

\item When the mass accretion rate is higher than the stability
line of steady hydrogen burning $\dot M_{\rm acc} > \dot M_{\rm stable}$,
we can make a forced nova with no mass ejection
by manipulating stop/restart of mass accretion. 
This will be explained in Section \ref{natural.forced}. 
\end{enumerate}

Among a huge number of galactic novae ($Z \sim 0.02)$ only one 
(PU Vul) is known to be the object without optically thick winds. 
ASASSN-16oh is an SMC object and could be in a low $Z$ environment. 
With a lower $Z$, the wind acceleration is weaker because the
lower iron abundance results in a smaller peak in the OPAL opacity.  
\citet{kat13hh} studied starts/ends of winds in various populations 
based on their steady-state approach. This approximation may not be good in 
the rising phase of strong nova outbursts. In this work, we study 
occurrence of winds using a stellar evolution code \citep{kat17sha}.

\subsection{Method and Model Parameters}

Our numerical code is the same as that described in \citet{kat17sha}. 
When the hydrogen-rich envelope of the WD expands after hydrogen ignites,
the wind mass-loss often occurs. 
Henyey-type codes, widely used in stellar evolution calculation, have 
difficulties in calculation of extended stages of nova outbursts. 
This led many authors assume some mass-loss in their numerical codes to 
continue calculation, but these mass-loss rates are not based on reliable 
acceleration mechanism of nova winds \citep{kat17palermo}. 
\citet{kat17sha} introduced how to obtain the mass-loss rate consistent 
with optically thick winds \citep{kat94h}. This method is to connect
an interior structure with an outer wind solution, and 
needs many iterations and human time until we obtain the final 
value of mass-loss rate. 
In the present paper, we focus our calculation on no-mass-ejection novae. 
Therefore, without such an iteration process, 
we simply assume, if needed, a tentative mass-loss rate 
described by Equation (1) of \citet{kat17sha}.

We assume Population II composition for accreted matter, i.e., 
$X=0.75$, $Y=0.249$, and $Z=0.001$,
where $X$ is the hydrogen, $Y$ is the helium, and $Z$ is the heavy 
element content in mass weight. This corresponds to a typical metallicity
of the SMC ([Fe/H]=$-1.25 \pm 0.01$) \citep{chi09}. 

\begin{figure}
\epsscale{1.1}
\plotone{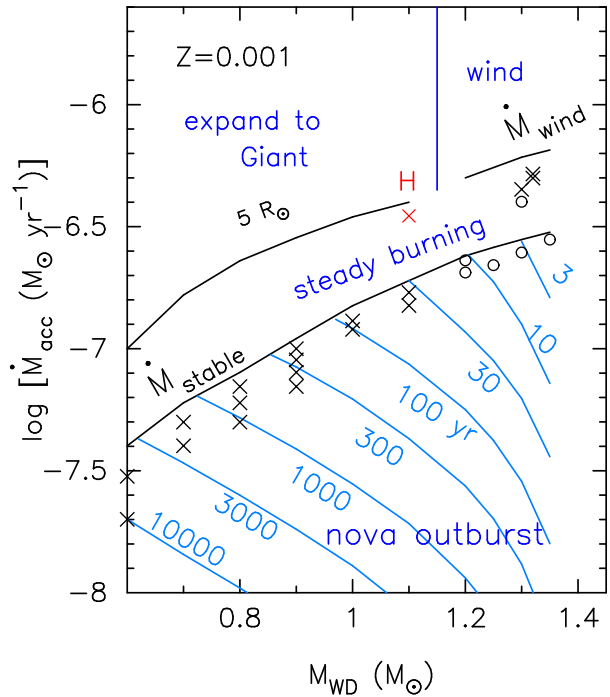}
\caption{The responses of WDs in the mass accretion rate
versus WD mass diagram. 
Below the stability line labeled $\dot M_{\rm stable}$,
hydrogen shell-burning is thermally unstable 
and the WD experiences repeated shell flashes. 
Above the line, hydrogen burning is stable and no repeated 
shell flashes occur unless the mass accretion rate is manipulated 
by some mechanisms (forced novae).  
The crosses denote the models in which no mass-ejection occurs during
the outburst while the open circles correspond to the models with mass-loss. 
The upper solid line labeled $5~R_\sun$ shows the place at which 
the photospheric radius reaches $5~R_\sun$.  Above the line, 
the photosphere could be larger than the binary separation and 
a common envelope evolution may occur. 
In the region right-side to the vertical blue line, optically thick winds are 
accelerated and the binary may experience an accretion wind evolution, 
avoiding a common envelope evolution. 
The red cross labeled ``H'' is the Hillman et al.'s (2019) model, 
which is above the stability line, and their model is intrinsically
a forced nova (see Section \ref{hillmanmodel}). 
\label{nomotoD}}
\end{figure}

The WD mass and mass accretion rate of our models 
are plotted in Figure \ref{nomotoD},
which is the diagram of WD response
for various WD masses and mass accretion rates.
Below the stability line of steady hydrogen burning
(lower black line labeled $\dot M_{\rm stable}$),  
hydrogen shell burning is unstable and the accreting WDs experience 
periodic shell flashes, i.e., nova outbursts. 
The equi-recurrence period of shell flashes is 
depicted by the cyan-blue lines, with the period beside each line. 
The crosses represent no-mass-ejection models while the open circles
depict models in which we assume mass-loss
otherwise our calculation does not converge. 

Above the stability line, the photospheric radius is larger
for a larger mass accretion rate.
In the region left to the vertical blue line
($M_{\rm WD} \lesssim 1.15 ~M_\sun$),
the envelope eventually expands to giant size without mass-loss.
The optically thick wind mass-loss occurs only in the region 
above the upper horizontal black line and right to the vertical blue line,
i.e., very massive WDs ($M_{\rm WD} \gtrsim 1.15 ~M_\sun$)
and high mass accretion rates ($\dot M_{\rm acc} \gtrsim 6\times 10^{-7}
M_\sun$ yr$^{-1}$).

\subsection{Forced Novae and Natural Novae}
\label{natural.forced}

\begin{figure*}
\epsscale{0.65}
\plotone{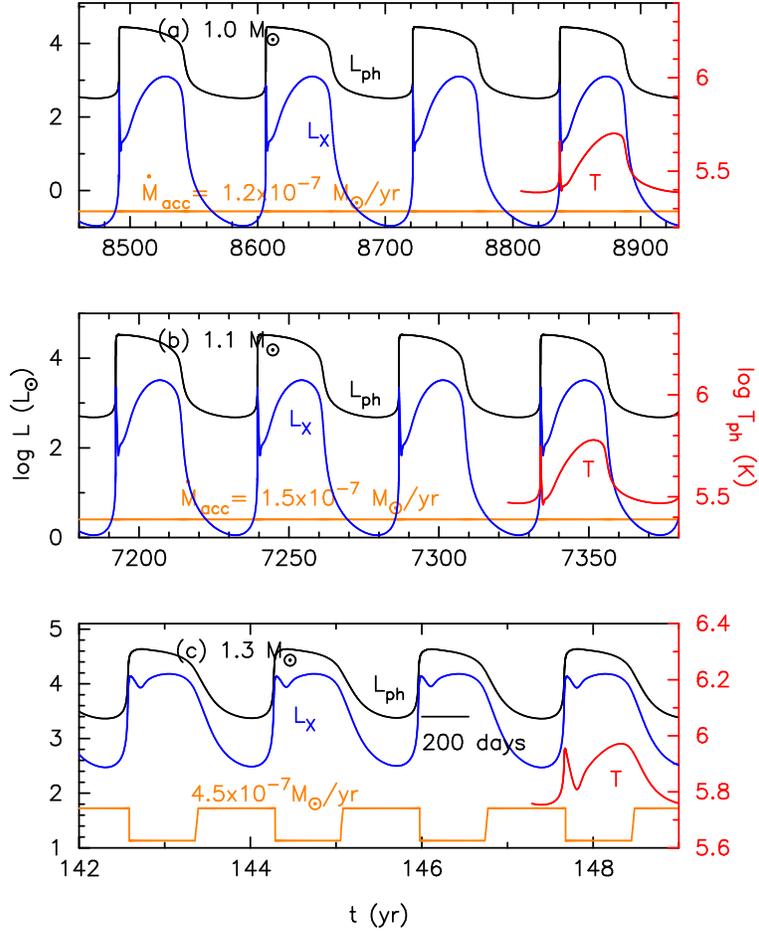}
\caption{Our models without mass-ejection are shown for several flashes:
the photospheric luminosity, $L_{\rm ph}$ (black line); 
supersoft X-ray ($0.3-1.0$ keV) luminosity, $L_{\rm X}$ 
(blue line); photospheric temperature, $T_{\rm ph}$ (red line:  
the last cycle only); and mass accretion rate, $\dot M_{\rm acc}$ 
(orange line).   The chemical composition of accreted matter 
is assumed to be $X=0.75$, $Y=0.249$, and $Z=0.001$. 
(a) A $1.0~M_\sun$ WD with 
$\dot M_{\rm acc}=1.2 \times 10^{-7}~M_\sun$ yr$^{-1}$. 
(b) A $1.1~M_\sun$ WD with 
$\dot M_{\rm acc}=1.5 \times 10^{-7}~M_\sun$ yr$^{-1}$.
(c) A $1.3~M_\sun$ WD with 
$\dot M_{\rm acc}=4.5 \times 10^{-7}~M_\sun$ yr$^{-1}$. 
The models in panels (a) and (b) are natural novae.
We do not stop the mass accretion in the outburst phase. 
The model in panel (c) is a forced nova
in which we stop the mass accretion 
($\dot M_{\rm acc}=0.0$) and then restart it 
($\dot M_{\rm acc}=4.5 \times 10^{-7}~M_\sun$ yr$^{-1}$) as denoted
by the orange line labeled $4.5\times 10^{-7}~M_\sun$/yr. 
\label{light}}
\end{figure*}

Figure \ref{light} shows the last four cycles of our three 
models.  No mass-ejection occurs in any of the models in this figure. 
The first two are close to but below the stability line, i.e., 
(a) $M_{\rm WD}= 1.0~M_\sun$ with $\dot M_{\rm acc}=1.2 \times 10^{-7}
M_\sun$ yr$^{-1}$ and (b) $M_{\rm WD}= 1.1~M_\sun$ with $\dot M_{\rm acc}=
1.5 \times 10^{-7}~M_\sun$ yr$^{-1}$. The WD always accretes matter. The 
accretion rate is smaller than the nuclear burning rate during  
the flash ($\dot M_{\rm acc} < \dot M_{\rm nuc}$), then the envelope mass 
gradually decreases ($\dot M_{\rm env}= \dot M_{\rm acc} - \dot M_{\rm nuc}
< 0$) to below the minimum envelope mass for hydrogen burning. 
Thus, hydrogen burning eventually ends. 
We call the novae that occur below the stability line ``natural nova'' 
to distinguish them from the ``forced novae'' that occur 
above the stability line \citep{hac16}. 

Figure \ref{light}(c) shows an example of forced novae, i.e., 
$M_{\rm WD}= 1.3~M_\sun$ with $\dot M_{\rm acc}=4.5 
\times 10^{-7}~M_\sun$yr$^{-1}$, being located above the stability line. 
In this region the nuclear burning rate is balanced to the accretion rate 
($\dot M_{\rm env}= \dot M_{\rm acc} - \dot M_{\rm nuc} = 0$). 
The decrease in the envelope mass is supplied by the mass accretion. 
Thus, if we continue the mass accretion, hydrogen-shell burning
will never end.  If we stop the mass accretion, the envelope mass
decreases by nuclear burning and the shell-flash eventually ends.
If we restart the mass accretion after some time later, 
we obtain the next shell flash. 
In this way, we have multi-cycle shell flashes. 
Figure \ref{light}(c) gives an example of the periodic change 
of the mass accretion rate. If we stop/restart the mass accretion 
in our model calculation, we obtain successive shell flashes. 

Figure \ref{light} compares characteristic properties of these 
three models: the photospheric luminosity $L_{\rm ph}$, 
photospheric temperature $T_{\rm ph}$, and supersoft X-ray light curve 
$L_{\rm X}$ (0.3 - 1.0 keV) calculated from $L_{\rm ph}$ and $T_{\rm ph}$,
using a blackbody approximation.
The photospheric temperature is shown only in the last cycle
to avoid complication of lines. 
As the photospheric temperature is always high 
($\log T_{\rm ph}$ (K) $> 5.4$), 
most of photons are emitted in the far-UV and supersoft X-ray bands. 
As expected, they are all periodic supersoft X-ray sources. 

\section{Search for novae having a short X-ray duration with no-mass-ejection}
\label{z001}

\subsection{Natural Novae} 

We searched shell-flash models for no-mass-ejection. 
Below the stability line, i.e., in the natural novae, we found 
no-mass-ejection models only in low mass WDs as shown in Figure \ref{nomotoD}. 
In more massive WDs ($M_{\rm WD} \gtrsim 1.1 ~M_\sun$), 
shell flashes are stronger and mass ejection always occurs. 
The border of with/without mass ejection is about 
$M_{\rm WD}= 1.0 - 1.1 ~M_\sun$ as in Figure \ref{nomotoD}.  
This is consistent with the study of occurrence condition of 
winds for metal-poor novae \citep{kat13hh}. 

Far below the stability line, i.e., with lower mass-accretion rates, 
the shell flash becomes much stronger and 
we have to assume large mass-loss rates during the outburst. 

The SSS duration is longer in lower mass WDs. 
In the $1.0$ and $1.1~M_\sun$ WD models, 
the massive limit of no-mass-ejection models has 
the SSS duration much longer than 200 days, being not consistent with
the ASASSN-16oh observation (see Figure \ref{light}(c)).  
Thus, we conclude that there is no natural nova corresponding 
to ASASSN-16oh.

\subsection{Forced Novae}

Above the stability line, i.e., in forced novae, 
the recurrence period can be controlled
by manipulating the re-starting time of accretion. 
If we delay the re-starting time the next outburst occurs 
later even if both the WD mass and 
mass-accretion rate are the same \citep{kat14shn, hac16}. 
The shell flash becomes stronger because the WD is cooled down 
during the delay.

As the recurrence period of forced novae can be controlled, we first search  
nova sequences for a short X-ray duration regardless of recurrence period.  
The outburst duration is shorter in more massive WDs. 
We found that a WD more massive than $1.32 ~M_\sun$ can have
an X-ray duration shorter than 200 days. In WDs more massive than 
$1.32~M_\sun$, however, we have to assume mass loss to continue numerical 
calculation.  Thus, the $1.32 ~M_\sun$ WD is 
the upper mass limit in our calculation that fulfills a short X-ray duration
($\lesssim 200$ days) and no-mass-ejection.
In the next subsection we describe our $1.32 ~M_\sun$ WD model in detail.

\begin{figure*}
\epsscale{0.65}
\plotone{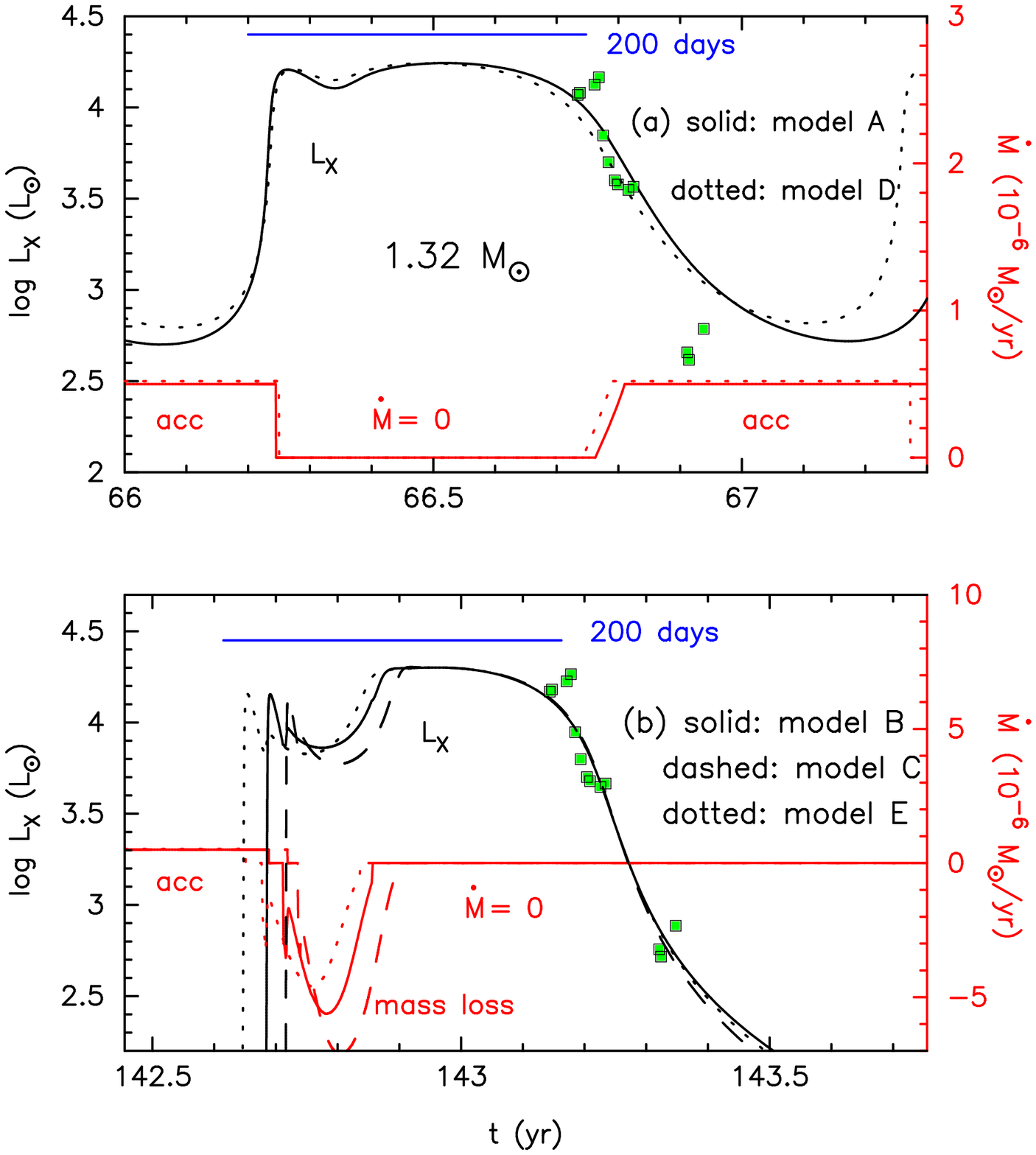}
\caption{Close up view of $L_{\rm X}$ (0.3 - 1 keV) 
in our forced nova model of a $1.32~M_\sun$ WD.
We added the {\it Swift} data for comparison (filled green square
outlined by black line). 
(a) Model A: $\dot M_{\rm acc}=5 \times 10^{-7}~M_\sun$ yr$^{-1}$ 
(solid line) and Model D: 
$\dot M_{\rm acc}=5.2 \times 10^{-7}~M_\sun$ yr$^{-1}$ (dotted line). 
We stop the mass accretion in the bright phase. Wind mass-loss does not  
occur. (b) Same as the models in panel (a) but with a delay  
of mass accretion.  Model B: solid line. Model C: dashed line.  
Model E: dotted line. 
The accretion restarts 2.6 yr after the $L_{\rm nuc}$ maximum 
in Model B, while 5.3 yr in Model C and 2.8 yr in Model E.   
The assumed mass-loss rates are depicted by the red lines. 
\label{light.m132}}
\end{figure*}

\subsection{$1.32 ~M_\sun$ Forced Nova Model}

Table \ref{table_models} summarizes shell flash properties of 
our five $1.32 ~M_\sun$ models.  It lists the WD mass ($M_{\rm WD}$),
metallicity ($Z$), mass accretion rate ($\dot M_{\rm acc}$),
restarting time ($t_{\rm restart}$) of accretion after previous
shell flash occurs, i.e., time from the epoch 
when the maximum nuclear luminosity is attained, 
adopting numerical mass loss or not (yes or no),
recurrence period ($t_{\rm rec}$), maximum nuclear luminosity
($L_{\rm nuc}^{\rm max}$), 
amount of accreted matter between two flashes ($M_{\rm acc}$), 
mass retention efficiency, i.e., ratio of the mass of ash helium 
and accreted matter  ($\eta$), and mean mass increasing rate of the WD 
(a CO core with a He layer) ($\dot M_{\rm CO}$). 
The properties of Models A and D are almost the same because their mass 
accretion rates are not so different.  Also Models B and E are similar. 
The maximum nuclear luminosity $L_{\rm nuc}^{\rm max}$ 
represents the flash strength. 
The flash is stronger for a smaller mass accretion rate, and 
a later re-starting time of accretion.

Figure \ref{light.m132}(a) shows two models of no-mass-ejection, 
Models A (solid lines) and D (dotted lines). 
As the mass-accretion rates are almost the same, 
the light curves are very similar. The next outburst 
begins a bit later in the smaller mass-accretion model 
(Model A). 
The duration of the X-ray light curves are about 200 days
in both models. 

Figure \ref{light.m132}(b) shows Models B, C and E, models  
with a later restarting time of mass-accretion.  In these cases,
the X-ray light curve decays faster than the models in panel (a). 
The recurrence period becomes longer to 4 years in Model B and E, 
and 7 years in Model C. 
In our calculation code, we cannot continue calculation 
in the expanded stage unless we adopt numerical mass-loss scheme,
the rate of which is shown by the red line in the figure.
As shown in Models B and C, we need to assume a larger mass-loss rate  
for a larger restarting time of mass-accretion. 
This does not directly mean that the mass-loss actually occurs
because we did not adopt iteration process to include realistic
wind solutions as we did in \citet{kat17sha}.
Even if a real mass-loss occurs, the acceleration is very weak in such
a low $Z$ environment and the wind velocity would be as small as
$\sim 100$ km~s$^{-1}$ \citep{kat99}, which is comparable with
the FWHM of \ion{He}{2} line in the optical spectra \citep{mac19}.

The wind durations of Models B, C, and E are 55--58 days as in Figure
\ref{light.m132}(b).  The winds stopped 84--106 days before the optical
detection of ASASSN-16oh.  The wind velocity is as small as 
100 km~s$^{-1}$ or smaller, much smaller than the escape velocity.
Therefore, the escaping matter would soon fall back to the equatorial plane
and stay as circumbinary matter or merge into the accretion disk.
We suppose that the escaping matter (wind) from the WD may not
contribute to the \ion{He}{2} line at the detection of ASASSN-16oh.

Due to numerical difficulties, we did not calculate other models
with similar parameters. We think that the 1.32 $M_\sun$ models
are close to the one that is consistent with the observation of ASASSN-16oh.

We show the X-ray count rate of ASASSN-16oh to the decay 
phase of these theoretical light curves. 
In this fitting we regarded that the hydrogen shell flash started 200 days 
before the X-ray detection. The WD was always bright in the X-ray band 
during this 200 days, but unfortunately, we had no chance of detection
until the very later phase.   
We further assume that a massive mass-inflow started 
shortly before the shell flash started. This mass-inflow may be 
associated with a dwarf nova outburst. 
The dwarf nova outburst developed from inside to out of the disk, 
which results in a slow rise of the $V/I$ light curves. 
The $V/I$ brightnesses could be brighter than normal dwarf nova outbursts 
because the disk is irradiated by the hot WD.  
irrespective of whether a dwarf nova triggered the nova outburst 
or a nova outburst triggered a dwarf nova outburst, 
this picture roughly explains observational 
properties of ASASSN-16oh discussed in Section \ref{sec_obs}. 

We also calculated a similar model, but for a lower metallicity, 
i.e., $Z=0.0001$, $M_{\rm WD}= 1.35 ~M_\sun$, and $\dot M_{\rm acc}= 5 \times
10^{-7} ~M_\sun$ yr$^{-1}$ with different restarting times
of accretion. These model properties are listed in Table \ref{table_models}.
Models F and G show no-mass-ejection, but Model H accompanies mass-loss.  
They have very similar X-ray light curves to those 
in Figure \ref{light.m132}. 
 
The models except Model C 
show shorter recurrence periods 
than 6 yr. This does not match the observation of ASASSN-16oh,
because at least a 6 yr quiescent phase is obvious prior to 
the outburst from the OGLE IV data \citep{mac19}. 

To summarize, searching for a model having a short X-ray 
duration ($< $ 200 days), no-mass-ejection, and long recurrence 
period ($> 6$ yr), we find that the 1.32 $M_\sun$ WD ($Z=0.001$) 
model is the closest one to the ASASSN-16oh observation.  

The mass retention efficiency is $\eta = 1.0$ for no-mass-ejection 
models and $\eta = 0.3-0.4$ for mass-ejection models. 
After one cycle of hydrogen shell flash, the ash helium accumulates
on the CO core.  This mass-increasing rate of CO core is listed
in the last column of Table \ref{table_models}, 
which amounts as large as $\dot M_{\rm CO} > 
2\times 10^{-7} M_\sun$ yr$^{-1}$ for no-mass-ejection models.

\begin{figure*}
\epsscale{0.65}
\plotone{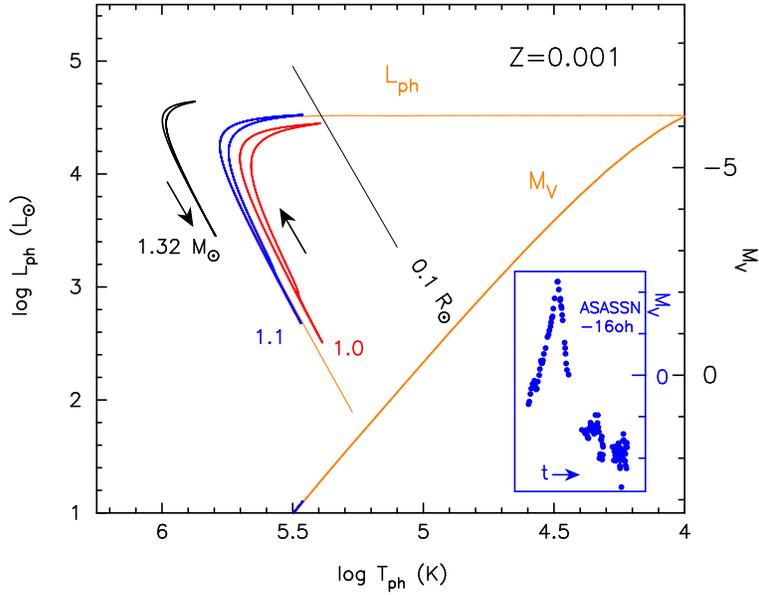}
\caption{One cycle of shell flashes in the HR diagram.  All the models 
have no mass-ejection.  From left to right.  
Black line: a $1.32~M_\sun$ WD with $\dot M_{\rm acc}=5 \times 
10^{-7}~M_\sun$ yr$^{-1}$ (Model A in Table \ref{table_models}). 
Blue line: a $1.1~M_\sun$ WD with $\dot M_{\rm acc}=
1.5 \times 10^{-7}~M_\sun$ yr$^{-1}$ in Figure \ref{light}(b).
Red line: a $1.0~M_\sun$ WD with $\dot M_{\rm acc}=
1.2 \times 10^{-7}~M_\sun$ yr$^{-1}$ in Figure \ref{light}(a).
The black arrows indicate the direction of evolution. 
The thin orange line shows the track of sequence of 
static solutions that represents a theoretical nova 
outburst on a $1.1~M_\sun$ WD with a very large ignition mass. 
The thick orange line labeled $M_V$ is 
the absolute $V$ magnitude on the rightside ordinate,
calculated from the temperature ($T_{\rm eff}$) 
and luminosity ($L_{\rm ph}$) of the thin orange 
model in the upper (bright) branch.  
We also add $M_V$ of the upper branch in the $1.1 ~M_\sun$ 
model (blue line model), which appears in the very bottom region
on the thick orange line. The thin black line indicates the 
locus of $R_{\rm ph}=0.1 ~R_\sun$. 
The inset shows the $M_V$ light curve of ASASSN-16oh, 
the same data in Figure \ref{as16oh_vi_linear}, but 
shifted with $(m-M)_V=19.0$. The naked WD without an irradiated 
accretion disk is darker than $M_V > 3$, and
cannot reproduce the $V$ brightness of ASASSN-16oh.
\label{hr}}
\end{figure*}

\subsection{Shell Flashes in the HR Diagram}
\label{naked_wd}

It is instructive to show how shell flashes behave 
in the HR diagram.  Figure \ref{hr} shows shell-flash tracks in 
the HR diagram for the last cycle of our three models, 1.32 $M_\odot$ WD (black line), 
1.1 $M_\odot$ WD (blue line), and 1.0 $M_\odot$ WD (red line).  
These models are already shown  in Figures \ref{light.m132} (Model A), 
\ref{light}(b), and \ref{light}(a), respectively. 

As the shell flash proceeds, the WD starts from the bottom of the 
curve and brightens up and turns to the right in the HR diagram.
The maximum expansion of the WD photosphere depends on the 
ignition mass. In these models, the ignition masses are very 
small so their envelopes do not extend beyond $0.1~R_\sun$
(thin black line). 
This properties are already studied with hydrostatic approximation 
in \citet{kat09} for $Z=0.02$. 

The thin orange line labeled as $L_{\rm ph}$ shows a 
sequence of the hydro-static solutions that represents 
a theoretical nova outburst on a $1.1~M_\sun$ WD with a very 
large ignition mass of hydrogen envelope. 
Here we assume that the chemical composition of the envelope is
$X=0.75$, $Y=0.249$, and $Z=0.001$ and the original WD radius
underneath the hydrogen-rich envelope is 
$\log (R_{\rm WD}/R_\sun)=-2.104$ taken from the $1.1~M_\sun$ model 
with the mass accretion rate of $\dot M_{\rm acc}=1.5 \times 
10^{-7}~M_\sun$~yr$^{-1}$. 
It should be noted that the photospheric luminosity in the upper branch
(horizontal part of the thin orange line) is almost the same as the Eddington
luminosity.

The thick orange line labeled as $M_V$ denotes the absolute
$V$ magnitudes correspond to the sequence of the thin orange line.  
Here we have calculated the absolute $V$ magnitude $M_V$ from 
$T_{\rm ph}$ and $L_{\rm ph}$ with the blackbody approximation.
The contributions from the accretion disk and companion star are 
not included. 
This figure is similar to Figure 15 of \citet{ibe82}.
Even if the $L_{\rm ph}$ is constant, the $M_V$ is smaller (brighter)
for a lower $T_{\rm ph}$ because of different bolometric correction. 
In our models, the photospheric temperature 
does not decrease much ($\log T_{\rm ph}$ (K) $> 5.4$),
so the $V$ brightness is rather faint. 
For example, our 1.1 $~M_\sun$ WD model (blue line) reaches 
the maximum at $M_{\rm V}=3$ as shown by the blue part 
on the lower orange line labeled $M_V$. 

The inset shows the absolute $V$ magnitudes of ASASSN-16oh. 
Its maximum brightness reaches $M_V \sim -2.3$,
which is much brighter than the naked WD. 
It is obvious that there is another source in the $V/I$ 
brightnesses of ASASSN-16oh.

\begin{figure*}
\epsscale{0.65}
\plotone{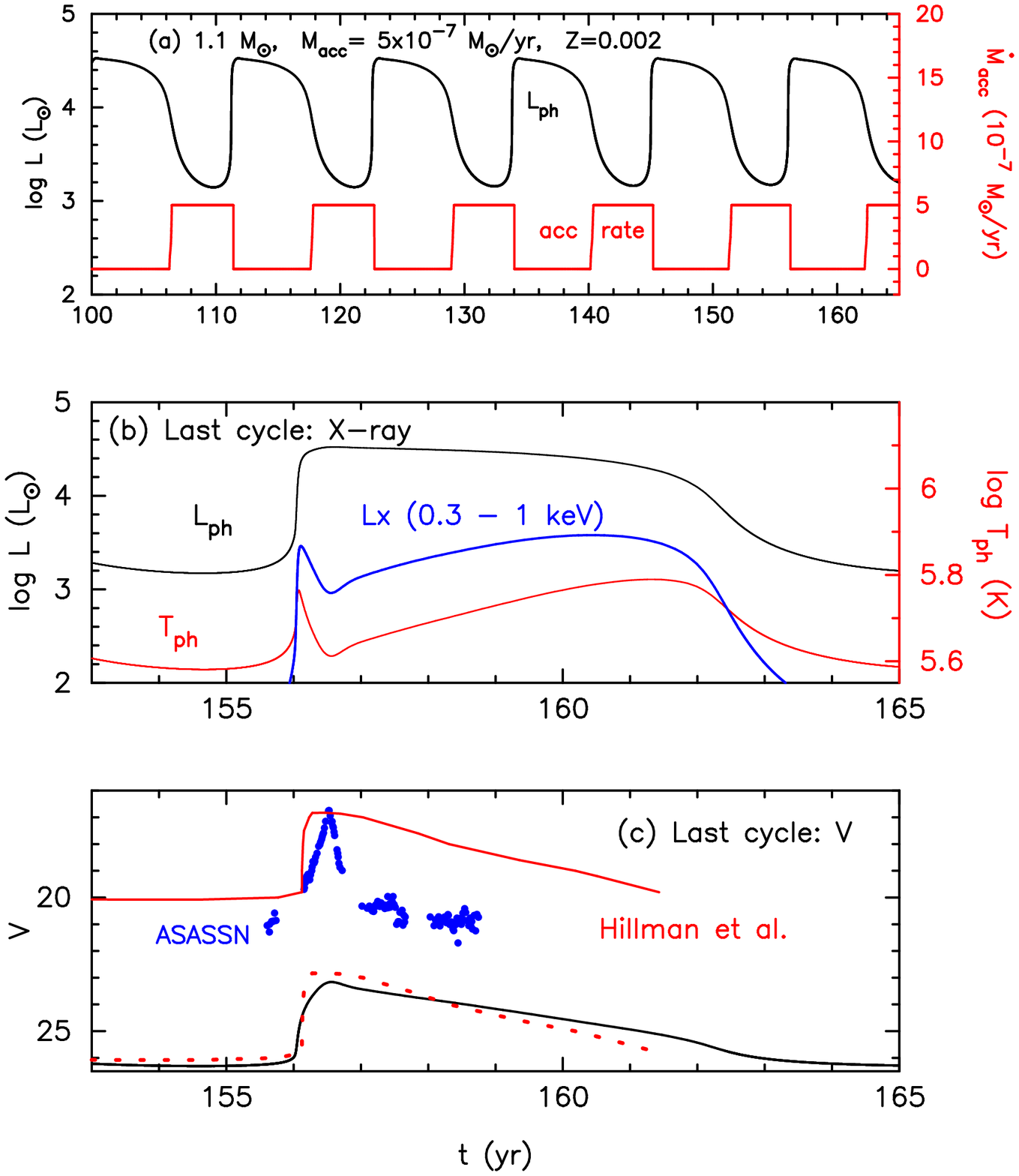}
\caption{ A shell flash model on a $1.1~M_\sun$ WD with 
the mass accretion rate of $\dot M_{\rm acc}=5\times 10^{-7}~M_\sun$ yr$^{-1}$ 
for the chemical composition of $X=0.70$, $Y=0.298$,
and $Z=0.002$. These parameters are taken from \citet{hil19}.
(a) The photospheric luminosity (black line) and manipulated 
mass accretion rate (red line) 
that we adopted to mimic Hillman et al.'s model.
(b) Close up view of the last cycle in panel (a). The photospheric 
luminosity (black line), photospheric temperature (red line), and 
supersoft X-ray luminosity ($0.3 - 1.0$ keV: blue line). 
(c) The model $V$ light curve (black line) calculated 
from the luminosity and temperature shown in panel (b).
Here, we adopt $\mu_V\equiv (m-M)_V=19.0$ for ASASSN-16oh. 
The estimated $V$ band light curve of ASASSN-16oh 
(in Figure \ref{as16oh_vi_linear}) is added 
in the same timescale for comparison (blue dots).  
ASASSN-16oh is much brighter, by $\sim$ 6 mag, than our theoretical model. 
We also add the $V$ light curve (solid red line)
of Hillman et al.'s model \citep[taken from Figure 2 of][]{hil19}.
This line agrees well with our model $V$ light curve 
if we shift it down by 6 mag (dotted red line).
\label{light.hillman}}
\end{figure*}

\section{Comparison with Hillman et al.'s (2019) calculation}
\label{hillmanmodel}

\citet{hil19} presented a nova model with no mass-ejection 
for ASASSN-16oh. This model is based on the idea 
that both the $V/I$ photons and X-rays are emitted from 
the hot WD surface. Their best fit $I$ light curves 
are obtained for $M_{\rm WD}=1.1~M_\sun$ and
$\dot M_{\rm acc}= (3.5 - 5) \times10^{-7}~M_\sun$ yr$^{-1}$. 

In Section \ref{z001}, however, we searched for possible nova models of
ASASSN-16oh and obtained the WD mass to be $M_{\rm WD} \ga 1.32~M_\sun$,
much more massive than Hillman et al.'s.   As the mass accretion rate
of $\dot M_{\rm acc}= 5 \times10^{-7}~M_\sun$ yr$^{-1}$ is above 
the stability line, we need some on/off-switch mechanism of mass accretion, 
otherwise we have steady-state burning \citep[see Figure 1 of][]{hac16}. 
In other words, a forced nova does not occur in nature without some 
mechanism of on/off switch. In our model this on-switch is 
a mass accretion during a dwarf nova outburst. 
As shown below, Hillman et al.'s model is a forced nova with manipulated 
mass-accretion in their computer code, although no description about 
this on/off-switch is given.

We calculated a shell flash model with the same parameters 
as Hillman et al.'s, i.e., $M_{\rm WD}=1.1~M_\sun$, 
$\dot M_{\rm acc}= 5 \times10^{-7}~M_\sun$ yr$^{-1}$, 
and $X=0.7$, $Y=0.298$, and $Z=0.002$. 
As shown in Figure \ref{light.hillman}(a) 
we have successive hydrogen shell flashes only if we 
manipulated the mass-accretion as indicated by the red line. 
The photospheric luminosity $L_{\rm ph}$, temperature 
$T_{\rm ph}$, and supersoft X-ray luminosity $L_{\rm X}$ in the 
last cycle of our calculation are shown in Figure \ref{light.hillman}(b). 
Figure \ref{light.hillman}(c) shows the $V$ light-curve of our model. 
The $M_{\rm V}$ magnitudes are calculated from $L_{\rm ph}$ and 
$T_{\rm ph}$ with the $V$ band response function and
converted to the $V$ magnitudes with the distance modulus of 
$\mu_V \equiv (m-M)_V = 19.0$ in $V$ band for ASASSN-16oh.

Figure \ref{light.hillman}(c) also shows the $V$ band light curve 
of ASASSN-16oh (blue dots) in comparison with our WD model. 
Our model light curve (black line) is much fainter, 
by 6 mag at the peak, than ASASSN-16oh.
The solid red line indicates the $V$ light curve calculated by 
\citet[][taken from their Figure 2]{hil19}. 
The peak magnitude is comparable to that of ASASSN-16oh, 
but decays much slower. 
The $V$ light curve agrees well with our light curve if we assume $(m-M)_V=13$ 
as shown by the dotted red line. 
We suspect that Hillman et al. might wrongly adopt $(m-M)_V=13$ 
(rather than 19) for the distance modulus in $V$ band of ASASSN-16oh.

To summarize, we find that Hillman et al.'s light curve
is fainter than ASASSN-16oh by 6 mag and decays 10 times slower.

\section{Comparison with Maccarone et al.'s model} \label{sec_maccarone}

Our shell flash model and Maccarone et al.'s (2019) spreading layer model 
predict qualitatively similar observational properties. 
Both models predict supersoft X-rays of blackbody spectrum 
from optically thick matter on a massive WD. 
Both models are based on a dwarf nova outburst,
in which the accretion disk and companion star is
more or less irradiated by X-rays from the WD. 

The main difference is in the emitting area and luminosity of X-ray
radiation, corresponding to the different energy sources. 
In our shell flash model nuclear burning is the energy source 
and the whole WD surface is bright. 
In the spreading layer model, the gravitational energy 
is released in the inner edge of the accretion disk, and 
the X-ray emitting region is concentrated on the equatorial belt. 
These differences result in different observational properties which 
we discuss in the following subsections.

\subsection{Total Flux and Binary Inclination of ASASSN-16oh}

In the spreading layer model, the maximum luminosity 
is roughly estimated from gravitational energy release, 

\begin{equation}
L^{\rm acc}= {G M_{\rm WD} {\dot M_{\rm acc}}\over 2 R_{\rm WD}}.
\label{eq.acc}
\end{equation}
For a 1.3~$M_\sun$ WD with $\dot M_{\rm acc}=3\times 10^{-7}M_\sun$~yr$^{-1}$ 
and $R_{\rm WD}=3 \times 10^8$ cm, we have 
$L^{\rm acc}= 5.2\times 10^{36}$erg~s$^{-1}$. 
On the other hand, our 1.32~$M_\sun$ WD model in Figure \ref{light.m132} 
emits $L_{\rm ph}=1.2 \times 10^{38}$ erg~s$^{-1}$, 
23 times larger than the spreading layer emission. 

As introduced in Section \ref{sec_obs}, \citet{mac19} estimated 
X-ray luminosity from Chandra spectrum of 
ASASSN-16oh to be $6.7 \times 10^{36}$ erg~s$^{-1}$. 
This value is consistent with the gravitational energy release 
from the spreading layer.  
\citet{hil19} also obtained the unabsorbed X-ray flux to be 
$(4.3 - 6.4) \times 10^{36}$~erg~s$^{-1}$. 
They adopted a thermonuclear runaway model, which would emit an 
X-ray luminosity of about $10^{38}$~erg~s$^{-1}$, so they explained the observed
small X-ray flux by occultation of the WD surface like in U Sco. 
We also assume the occultation of the WD surface by the inflated 
disk edge, that is, we can not directly observe the WD emission 
but detect scattered photons which are much fewer than 
the original WD emission. 
It could occur if the binary inclination is high enough like in U~Sco.

\citet{mac19} estimated the inclination angle of the binary
from the span of the velocity variation in the \ion{He}{2} line 
(75 km~s$^{-1}$).  They assumed that the \ion{He}{2} line
comes from the central part of the accretion disk.
For a binary consisting of a $1.3~M_\sun$ WD and 0.7 $M_\sun$ companion, 
they obtained the binary inclination 
from Keplerian law to be $i=12\arcdeg$ for a orbital 
period of $P_{\rm orb}=1$ day, and $i=27\arcdeg$ for $P_{\rm orb}=10$ days.
This value should be replaced by $i=23\arcdeg$ and $i= 59\arcdeg$,
respectively, because the authors misplaced the masses of the WD and donor 
in their calculation. 

The above argument on the inclination angle 
of the binary should be modified if the \ion{He}{2} emission line is associated to 
the hot spot or irradiated companion. 
If the emitting region is close to the center of 
mass of the binary, the small line velocities are obtained 
even in a high inclination binary. 
Also the line velocity span may not always represent 
the radial velocity amplitude because of the sparse observation. 

The accretion disk around a very hot WD may not be geometrically thin
but could be inflated. The disk surface absorbs 
a part of the supersoft X-ray flux from the WD to become hot 
and emit higher energy photons. 
This irradiation and reprocess make the accretion disk 
inflated and brighter. 
\citet{pop96} assumed the $\tan^{-1} (z/\varpi)=15\arcdeg$ inflation in the 
UV-optical spectrum model for CAL 83 and RX J0513.9-6951. 
\citet{sch97} adopted an inflated disk shape of the angle 
up to $\tan^{-1} (z/\varpi)=24\arcdeg$ 
in the light curve model of the LMC supersoft X-ray source CAL 87. 
\citet{hkkm00} adopted $\tan^{-1} (z/\varpi)=17\arcdeg$ 
at the edge of the disk 
in their light curve model of U Sco 1999 outburst.  
Here, $z$ is the height of the disk surface from the equatorial plane 
and $\varpi$ is the distance from the central WD.

The inclination angle of the binary could be one of the 
key parameters to distinguish the two models.  
If ASASSN-16oh is an eclipsing binary, it strongly 
support the shell flash model rather than the spreading layer model. 
If the binary is face on, the faint X-ray flux of ASASSN-16oh 
is consistent with the spreading layer model, while   
the shell flash model needs some explanation for the faint X-ray flux. 
The present status indicates that the observational clues
are not sufficient to draw the definite conclusion.

\subsection{UV/optical SED} 
\label{sec_uvdisk}

Figure \ref{uvot} shows the spectral energy distribution (SED) of 
ASASSN-16oh \citep{mac19}. This spectrum resembles 
those of RX J0513.9-6951 taken by \citet{pak93} 
in the optical high state.  
RX J0513.9-6951 is a binary consisting of a massive WD that undergoes 
steady hydrogen burning, irradiated accretion disk, 
and irradiated companion. \citet{hac03RXJ} made light curve models 
(solid magenta line in Figure \ref{vmagfit_u_sco_rxj0513_as16oh}(b)) 
in which the irradiated disk mainly contributes to the optical 
brightness. The resemblance with RX J0513.9-6951 
indicates the presence of bright irradiated accretion disk 
and companion in ASASSN-16oh. 

\citet{pop96} calculated a composite spectrum for LMC persistent 
supersoft X-ray sources (thick solid black line in Figure \ref{uvot}). 
Their model binary consists of a hydrogen burning WD of $1.2~M_\sun$,   
irradiated accretion disk with a radius of $1.9~R_\sun$,
and Roche-lobe-filling $2~M_\sun$ main-sequence companion. 
The orbital period is $P_{\rm orb}= 1$ day. 
The inflated accretion disk ($\tan^{-1} (z/\varpi)=15\arcdeg$
at its outer edge)  
is irradiated and brightened by 19 times 
at $\lambda = 1000$ \AA, and 4 times at $\lambda =6300$ \AA.  
As a results, the total flux, the summation of these three components,
reproduces the $F_\lambda \propto \lambda^{-2.33}$ law.
The WD does not contribute much in the UV/optical region,  
but the hydrogen burning WD is necessarily because 
its large luminosity is the source of irradiation. 

The composite spectrum by Popham and Di Stefano is fainter by 
$\Delta \log F_\lambda \sim 0.3$ than that of observed ASASSN-16oh data. They 
assumed the WD luminosity of $L_{\rm WD}=1.5 \times 10^{38}$ 
erg~sec$^{-1}$, which is comparable to our H-burning WD model 
($1.2 \times 10^{38}$erg~sec$^{-1}$). 
If we assume that ASASSN-16oh is a binary of a $1.3~M_\sun$ WD 
and a $0.7~M_\sun$ companion 
with $P_{\rm orb}= 5$ days, the binary separation $a$ is 
$((1.3+0.7)/(1.2 + 2))^{1/3}~5^{2/3}=2.5$ times larger 
than Popham and Di Stefano's model. If we scale up the binary size 
by a factor of 2.5, without changing the configuration,  
the irradiated area of the disk and companion become $2.5^2=6.25$ times larger. 
This makes the flux increase by $\Delta \log F_\lambda \sim \log 6.25=+0.80$. 
If we adopt the binary inclination of $80 \arcdeg$, instead of Popham 
and Di Stefano's $60 \arcdeg$, then, the flux decrease owing to inclination 
is calculated to be $\log( \cos 80/ \arcdeg \cos 60\arcdeg)=-0.46$. 
Then, the flux calculated by Popham and Di Stefano 
should be increased by $0.80-0.46= 0.34$ for our assumed binary. 
The resultant spectral energy distribution is very 
consistent with that of ASASSN-16oh obtained by \citet{mac19}.

In the spreading layer model, the irradiation effects 
is small because the X-ray luminosity is small. 
The main contributer in the UV/optical range is 
the dwarf nova outburst. In general, the absolute flux of a 
dwarf nova is smaller than those shown in Figure \ref{uvot} 
and the wavelength dependence was reported to be
$F_\lambda \propto \lambda^{-\Gamma}$, $\Gamma \sim 1.5-2.3$ 
in the UV/optical range \citep[e.g.,][]{par19}.

\subsection{X-ray light curve} \label{sec_Xray}

In our shell flash model the X-ray flux begins to decrease 
when hydrogen burning almost dies out and the burning 
zone temperature gradually decreases. 
The speed of cooling depends on the WD mass; a more massive 
WD cools faster. 

Figure \ref{vmagfit_u_sco_rxj0513_as16oh}(a) compares 
the light curves of ASASSN-16oh with those of U Sco 
in a normalized time scale.  
U Sco hosts a massive WD, $1.37~M_\odot$ \citep{hkkm00}, 
more massive than our $1.32~M_\odot$ for ASASSN-16oh,  
and then the cooling time is much shorter. 

When the X-ray flux decreases with time, the irradiation 
effects decrease so that the UV/optical fluxes from 
the irradiated disk and companion also become 
faint. Thus, both the X-ray and optical fluxes decrease 
at the same time. In this way, our shell flash model 
naturally explains the simultaneous behavior in 
the X-ray and optical fluxes along with those of U Sco as 
in Figure \ref{vmagfit_u_sco_rxj0513_as16oh}(a). 

In the spreading layer model, the optical flux 
decreases when the dwarf nova outburst approaches its 
end. As the X-ray flux is closely related to the mass-accretion rate 
onto the WD, the expected X-ray count rate decreases 
with decrease in the optical flux \citep[e.g. ][]{osa96}. 
It is, however, unknown whether a dwarf nova outburst could show 
the simultaneous decrease, like in U Sco, in both the X-ray and 
optical fluxes in the normalized timescale in 
Figure \ref{vmagfit_u_sco_rxj0513_as16oh}(a).

\subsection{Non-detection of supersoft X-rays in the quiescent phase 
of RS Oph}
\label{sec_rsoph}

RS Oph is a recurrent nova that hosts a massive WD with high 
mass-accretion rate ($1.35~M_\sun$ and $1.2\times 10^{-7}~M_\sun$ 
yr$^{-1}$: \citet{hac06b,hac01kb}). 
The WD mass and mass accretion rate are more or less similar
to ASASSN-16oh. If the X-ray emission of ASASSN-16oh is emitted
from the optically thick spreading layer, we may expect 
supersoft X-rays in the quiescent phase of RS Oph. 

\citet{nel11} observed quiescent phase of RS Oph with the Chandra and
XMM-Newton satellites, 537 days and 744 days after the nova outburst. 
They analyzed the spectrum and obtained the maximum allowable 
blackbody temperature for an optically thick component of 
the spreading layer to be 0.034 keV (=3.9 $\times 10^5$ K). 
The upper limit of unabsorbed luminosity is  
$(1.1-1.6)\times 10^{35} (d/1.6 {\rm ~kpc})^2$~erg~s$^{-1}$, 
where $d$ is the distance to the star.  
This upper limit is smaller than those detected in ASASSN-16oh by 
a factor of 40.  

The boundary layer between a WD and the inner edge of the accretion 
disk has been studied by various authors. 
\citep[][ and references therein]{pop95,pir04}. 
If the mass accretion rate is large, the boundary layer is 
geometrically thin, optically thick, and emits supersoft X-rays. 
When the accretion rate decreases, the boundary layer becomes 
optically thin, geometrically thick, and emits hard X-rays. 

\citet{pop95} calculated the mass accretion rate at 
this transition. For a $1.0~M_\sun$ WD, the critical mass-accretion 
rate is $\log \dot M_{\rm acc}~(M_\sun$~yr$^{-1})=-6.12$.  
This value depends on the model parameters, e.g., such as 
the WD rotation rate, viscosity parameter. 
They did not calculate for $M_{\rm WD} > 1.0~M_\odot$, but 
from the tendency of quick increase with the WD mass, 
the transition may occur at 
$\log \dot M~(M_\odot$~yr$^{-1})\sim -5$ 
for $1.3~M_\sun$. 
Below this rate, the boundary layer is optically 
thin and would emit hard X-rays. A small accretion rate 
may be the reason why the supersoft 
X-ray component was not detected in the quiescent phase of RS Oph.  

If the same argument is applied to ASASSN-16oh with an accretion 
rate of $3\times 10^{-7}~M_\odot$~yr$^{-1}$, 
the boundary layer should be optically thin and  
hard X-ray component had to be expected during the 
outburst. However, there are no hard X-ray component 
in the spectrum \citep{mac19}. 
The possible reasons for the absence of hard X-rays are:\\
(1) A hot optically thin extended boundary layer was formed 
and emitted hard X-rays. However, the X-rays were occulted 
by the inflated disk edge and undetected, 
or its flux was too small to be detected.
\\
(2) The mass accretion had already stopped before the period 
 of observation.

\subsection{Summary}

The resemblance of the $V-I$ color 
(Figure \ref{hr_diagram_asassn16oh_u_sco_outburst_vi}) and 
absolute magnitude $M_V$ 
(Figure \ref{vmagfit_u_sco_rxj0513_as16oh}) indicates 
that ASASSN-16oh is a binary system consisting of an hydrogen 
burning WD, irradiated accretion disk, and lobe-filling companion star. 
The spectral energy distribution $F_\lambda \propto \lambda^{-2.33}$
(Figure \ref{uvot}) can be explained by the sum of contributions from 
an accretion disk and a companion star which are irradiated by a WD 
in the luminous stage of a shell flash. 
Our shell flash model is consistent with these three properties.
We further show that the simultaneous declines in the X-ray and optical
are also explained by our hydrogen burning model (Sections \ref{sec_Xray})
and that our hydrogen burning model is consistent with undetected
hard X-ray flux even for the optically thin boundary layer case 
(Section \ref{sec_rsoph}). 

ASASSN-16oh is an atypical dwarf nova in its slow rise/decay 
and large amplitude. It is very helpful to see  
if similar shape dwarf novae show similar color and brightness
to ASASSN-16oh.

\begin{figure}
\epsscale{1.1}
\plotone{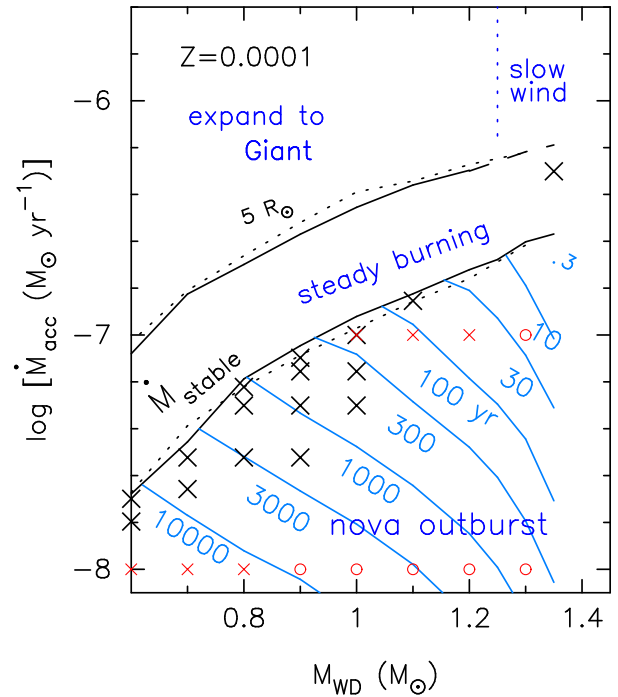}
\caption{ Same as Figure \ref{nomotoD}, but for Z=0.0001. 
The small red crosses, open red circles, and two dotted black lines 
are taken from \citet{che19}.  See the main text for more detail.
\label{nomotoDz0001}}
\end{figure}

\begin{deluxetable*}{lllllllllccl}
\tabletypesize{\scriptsize}
\tablecaption{Summary of forced Nova Models
\label{table_models}}
\tablewidth{0pt}
\tablehead{
\colhead{Model} &
\colhead{  } &
\colhead{$M_{\rm WD}$} &
\colhead{$Z^a$} &
\colhead{$\dot M_{\rm acc}$} &
\colhead{$t_{\rm restart}^b$}&
\colhead{Mass loss}&
\colhead{$t_{\rm rec}$} &
\colhead{$L_{\rm nuc}^{\rm max}$} &
\colhead{$M_{\rm acc}$} &
\colhead{$\eta$} &
\colhead{$\dot M_{\rm CO}$} \\
\colhead{}&
\colhead{}&
\colhead{ ($M_\odot$) } &
\colhead{  } &
\colhead{($10^{-7}M_\odot$ yr$^{-1}$)   } &
\colhead{ (yr)}&
\colhead{ }&
\colhead{ (yr)}&
\colhead{($ 10^5 L_\odot $) } &
\colhead{($10^{-7}M_\odot$)   } &
\colhead{}&
\colhead{($10^{-7}M_\odot$ yr$^{-1}$)   } 
}
\startdata
A & ... &1.32&0.001&5.0 &0.55 &no& 1.1   & 1.2  &2.9& 1.0&2.6 \\
B & ...   &1.32&0.001&5.0 &2.6  &yes& 4.2   & 10  &8.6& 0.34 &0.64\\
C & ...&1.32&0.001&5.0 &5.3 &yes& 7.3   & 13  &9.5 & 0.26&0.35 \\
D & ...&1.32&0.001&5.2 &0.53 &no& 1.0   & 1.0  &2.7 & 1.0&2.6 \\
E & ...  &1.32&0.001&5.2 &2.8  &yes& 4.2   & 9.5  &7.5 & 0.41&0.72 \\
F & ...&1.35&0.0001&5.0 &0.58&no& 1.1  &0.85 &2.7  & 1.0 &2.4\\
G & ...&1.35&0.0001&5.0 &0.80&no& 1.4  &1.2 &3.3  & 1.0 &2.3\\
H & ...   &1.35&0.0001&5.0 &1.5&yes& 2.5  &2.7   &5.0 & 0.75&1.5
\enddata
\tablenotetext{a}{$X=0.75$, $Y=1.0-X-Z$.}
\tablenotetext{b}{restarting time of mass-accretion 
since the $L_{\rm nuc}$ peak.}
%
\end{deluxetable*}

\section{Discussion}
\label{discussion}
\subsection{A Dwarf Nova Triggers a Nova Outburst?}

We examine the possibility if a dwarf nova outburst can trigger 
a nova outburst. 

An accreting WD will experience a nova outburst when the mass of
hydrogen-rich envelope reaches a critical value. 
Table \ref{table_models} list the amount of accreted matter 
to trigger hydrogen ignition.
They are a few to several times $10^{-7}~M_\sun$.
This ignition mass is smaller for a shorter recurrence period 
because the WD surface is hotter.

On the other hand, the mass inflow rate onto the WD is 
estimated in several dwarf novae \citep{kim18}.
Among them, V364~Lib shows a relatively large mass inflow rate of
$1 \times 10^{-7}~M_\sun$ yr$^{-1}$ \citep{kim18}.  
This object has an orbital period of $P_{\rm orb}= 0.70$ days, 
rising (decay) time of 10 (35) days, and small outburst 
amplitude of 1 mag.  As ASASSN-16oh has a much larger outburst  
amplitude and longer orbital period,  
we may expect a larger mass inflow rate.  If we assume the 
mass inflow rate of several times $10^{-7}~M_\sun$ yr$^{-1}$, 
the WD may accrete sufficient mass for hydrogen ignition. 
Our calculation assumed a periodic change of mass accretion, 
but in the real world, the mass inflow may occur irregularly. 
If WDs accrete sufficient mass through one or several times of 
mass-accretion events with a large mass-inflow rate,
a weak shell flash may occur similarly to ASASSN-16oh.

\subsection{Forced Novae}

Nova outbursts with high mass-accretion rates have been calculated by 
many authors \citep{pri95, yar05, ida13, hil15, hil19}. 
Forced nova phenomena are studied in detail by 
\citet{kat14shn} and by \citet{hac16} in which they pointed out that  
forced novae are already calculated 
without recognizing it because they automatically switched on/off
the mass accretion in their numerical codes. 
In the cases with very high mass-accretion rates,
characteristic properties are similar to those in Figure \ref{light},
i.e., no-mass-ejection and a short quiescent phase comparable to
the shell flash duration.

The forced novae are still theoretical objects because no mechanism
of switching-on/off is identified. 
If the ASASSN-16oh outburst is a shell-flash phenomenon triggered by
a dwarf nova, this is the first observational support for a forced nova.

In the forced novae, the mass accretion rate is very high by its definition, 
so the shell flash could be weak. 
Thus, the mass retention efficiency of the WD is very high. 
Especially with no-mass-ejection, the mass retention efficiency
of the WD is $\sim 100 \%$. 
Such an object is a candidate of Type Ia supernova progenitors. 
In this sense, ASASSN-16oh is a promising object.

\subsection{Novae in a Very Low Metallicity of $Z=0.0001$}

Nova outbursts in various metallicities were systematically 
studied by \citet{kat97,kat99} and by \citet{kat13hh}. 
The optically thick winds, the main mechanism of mass loss 
during the nova outbursts, are accelerated owing to 
the prominent iron peak in the radiative opacity 
at $\log T$ (K) $\sim 5.2$. For a lower metallicity, this Fe peak 
is small and almost disappears in $Z \lesssim 0.0004$. 
In such a case, another peak 
due to He ionization \citep[$\log T$ (K) $\sim 4.7$, 
see Figure 4 of][for opacities]{kat13hh} works 
as the driving source. 

\citet{kat13hh} showed that the Fe opacity peak works to 
drive the winds for a wide range of the WD masses ($M_{\rm WD} 
> 0.5~M_\sun$) for $Z=0.02$. For $Z=0.001$, it works 
only in massive WDs ($M_{\rm WD} > 1.05 ~M_\sun$) and 
the main driving source is replaced by the He opacity peak
in less massive WDs. 
For $Z \leq 0.0004$, only the He opacity peak works 
because the Fe peak disappears.
The acceleration by the He opacity peak may be 
weak because its peak is small.
Thus, we expect weak mass-ejection or slow expansion. 
Observational counter-parts of such phenomena 
are not identified yet, and then, 
it is not known if such a weak wind results in a nova 
outburst with efficient mass ejection from the binary.  

Figure \ref{nomotoDz0001} shows the stability line and recurrence 
periods for the composition of $X=0.75$ and $Z=0.0001$.  
In the same figure, we added the results by \citet{che19}. 
The small red crosses denote flashes with no-mass-ejection and 
the small open red circles are for those with mass-loss.   
Their results are consistent with ours. 

This plot also shows their stability line (lower dotted line) 
and their expansion line above which the WD envelope increases 
(upper dotted line).
These two lines are consistent with our $\dot M_{\rm stable}$ 
and $5~R_\sun$ lines.

\subsection{Metallicity Dependence of Stability Line}

The stability line has ever been calculated by many authors 
\citep[][for $Z=0.02$]{nom07, kat14shn, wol13, wol13b, ma13}, 
\citep[and][for $Z=0.0001$]{che19}. Our stability lines are 
lower than that of $X=0.7$ and $Z=0.02$ \citep{kat14shn} 
by a factor of 0.8 - 0.9 (vertically lower by $0.05 - 0.1$ dex)  
for $X=0.75$ and $Z=0.001$ in Figure \ref{nomotoD}, 
and by a factor of 0.6 - 0.8 (vertically lower by $0.1 - 0.2$ dex) 
for $X=0.75$ and $Z=0.0001$ in Figure \ref{nomotoDz0001}.

The stability line shows the minimum mass-accretion rate 
below which the hydrostatic envelope cannot keep the  
temperature high enough to support steady nuclear burning 
\citep[point B in Figure 1 of][]{kat14shn}. 
For the same WD mass and accretion rate, the envelope mass 
is larger for a lower $Z$ because of difference in the opacity. 
Thus, the temperature at the bottom 
of the envelope is also high. This makes the difference in 
the stability line.

\subsection{Metallicity Dependence of Recurrence Period}

Figure \ref{nomotoD} also shows the equi-recurrence period line.
The recurrence period is shorter for massive WDs and higher 
mass accretion rates. 

The equi-recurrence period lines were obtained for the solar composition 
($X=0.7$ and $Z=0.02$) by \citet{kat14shn} and by \citet{hac16}. 
Our equi-recurrence period lines are systematically longer than 
their equi-period lines by a factor of $\sim 1.4$ (shifted by 0.15 dex upward) 
in Figure \ref{nomotoD} (for $X=0.75$ and $Z=0.001$) 
and by a factor of $\sim 1.8$ (shifted by 0.25 dex upward) in 
Figure \ref{nomotoDz0001} (for $X=0.75$ and $Z=0.0001$).

For a lower $Z$, the ignition mass is larger because the lower opacity  
results in a lower blanket effect and it takes more time until 
a hydrogen ignition occurs at the bottom of the accreted matter. 
Thus, for the same WD mass and mass accretion rate, it takes 
a longer accretion time until the flash begins.

\section{Conclusions}
\label{conclusion}
Our main results are summarized as follows.

\begin{enumerate}
\item
We summarize observational properties of ASASSN-16oh from the view point
of an irradiated accretion disk model. 
The X-ray count rates are consistent with the scattered flux
from the obscured WD in the SSS phase of the recurrent nova U Sco. 
The $V-I$ color, peak $M_V$, and $F_\lambda\propto \lambda^{-2.33}$ law
are quite consistent with large irradiated accretion disks of U Sco
and RX J0513.9$-$6951. 
Thus, we may conclude that the supersoft X-ray fluxes 
originate from the hydrogen burning WD while the $V$ and $I$ band fluxes
come from the irradiated accretion disk. 

\item 
Supposing that a dwarf nova triggers a forced nova, we are able to obtain
the X-ray light curve consistent with ASASSN-16oh, i.e., a short X-ray
duration ($< 200$ days) and no mass-ejection. 
We found a $1.32~M_\sun$ WD model is the closest one for 
the low metallicity environment of $Z=0.001$. 
If it is the case that ASASSN-16oh is a dwarf-nova and it triggers
a nova, this is the first identified forced nova that occurs
above the stability line. 

\item
We have calculated a shell flash model with the same parameters 
as those in Hillman et al. (2019) and confirmed that their models 
are forced novae in which they manipulated on/off-switch of
the mass-accretion in their computer code. 
We have found that their light curve to fit with 
ASASSN-16oh are based on a too small distance modulus for the 
SMC. If we corrected, the peak and decay timescale are very different 
from those of ASASSN-16oh. 

\item 
The galactic nova PU~Vul was first observationally identified 
as a no-mass-ejection nova, but it occurred on a low mass WD 
($\sim 0.6~M_\sun$).    
If ASASSN-16oh is a thermonuclear runaway object like PU~Vul,
it is the second identified no-mass-ejection nova. 
No-mass-ejection novae hosting a very massive WD are a new type of SSSs
in low $Z$ environments.  They could be a candidate of Type Ia supernova 
progenitors because the mass retention efficiency is very high
($\eta \sim 100 \%$). 
 
\end{enumerate}

\acknowledgments
We thank the anonymous referee for useful comments that 
improved the manuscript.

%


\end{document}